# Receiving Exoplanet Messages with Gravitational-wave Carriers: The Gravitational-wave Radial Velocity Method

Casey McGrath[1,2,3]
[1] Center for Space Sciences and Technology, University of Maryland, Baltimore County, Baltimore, MD 21250, USA; casey.d.mcgrath@nasa.gov
[2] Center for Research and Exploration in Space Science and Technology II, NASA/GSFC, Greenbelt, MD 20771, USA
[3] Gravitational Astrophysics Lab, NASA Goddard Space Flight Center, Greenbelt, MD 20771, USA

## Abstract

Gravitational-wave (GW) radial velocity (RV) may offer a new method for detecting exoplanets, one that would primarily target planets orbiting GW-emitting, ultracompact, post-main-sequence stellar objects such as white dwarfs, neutron stars, and black holes. In this work we show that while similar to electromagnetic-based RV, this GW-based method depends on the "phase modulation index" (PMI), a parameter that is more sensitive to high-mass, long-period exoplanets. We investigate the potential of GW RV for GW sources in both the mid-frequency spectrum (where the upcoming LISA mission will operate) and the high-frequency spectrum (where the current LIGO–Virgo–KAGRA (LVK) observatories operate). Remarkably, we find that the PMI increases by a factor of $\sim\mathcal{O}(10^4 - 10^6)$ going from the mid-frequency to the high-frequency GW spectrum, which results in a greater redistribution of observed power due to the exoplanet's presence. By analyzing the power content of the observed GW signal, we derive analytic expressions to predict the GW carrier frequency power loss due to the exoplanet and the signal-to-noise ratio of the GW source and the exoplanet. We also explore the effects of the additional Doppler effect caused by the motion of the GW observatory and the potential for GW source parameter bias introduced by the presence of an undetected exoplanet. All of this helps us better understand where the GW RV method may work best in the exoplanet mass–period discovery space.

## 1. Introduction

As the field of gravitational-wave (GW) astrophysics has grown, so too have the ideas of discovery possibilities with these observatories. Primarily these observatories are designed to study the exotic and dead remnants of stars that have lived through their lives—black holes, neutron stars, and white dwarfs. But N. Seto (2008) described the possibility for using GW observations to detect exoplanets from a Doppler effect, envisioning the upcoming Laser Interferometer Space Antenna (LISA) mission as the instrument that could achieve such a discovery.[4]

The single most abundant type of GW source LISA anticipates discovering are double white dwarfs (DWDs) in the Milky Way. So the idea was that if an exoplanet orbited such a binary system, then its presence may leave a mark on the observed GW signal. T. Robson et al. (2018) studied and developed models for describing triple-body systems, envisioning the third body to be a stellar companion. N. Tamanini & C. Danielski (2019) and C. Danielski et al. (2019) then turned the attention to exoplanetary companions, developing the synergies of this new exoplanet discovery method with other methods and testing the detection possibilities. Numerous studies, cited throughout the course of this paper, have explored population studies, waveform modeling, detection prospects, and electromagnetic (EM) synergies for these types of exoplanet systems. An example of one such study, which extends the work presented specifically in this paper by investigating the detection and discovery synergy with a relatively new exoplanet detection method, infrared excess, is presented in C. McGrath et al. (2026).

The idea for detecting exoplanets (or other companions) around GW sources generally can apply to other missions beyond just LISA and other systems besides just DWDs (as examples, Y. Kang et al. 2021 compare detection prospects between both Taiji and LISA, and W.-L. Guo et al. 2025 consider binary neutron star (BNS) systems that could be found in the dHz GW spectrum). More generally, this is the idea for a new exoplanet detection method: the radial velocity (RV) modulation of a GW signal. Such a method would open up the intriguing possibility of future multimessenger exoplanet discovery.

*The RV method applied to GW observations.* If an exoplanet orbits a GW source, then it may induce a Doppler modulation on the observed GW signal. This would be the RV method for detecting an exoplanet, but applied to GW observations. This described Doppler modulation would be an *observed* effect—an alteration of the GW signal as seen by the observer. If, however, the orbiting body were sufficiently massive and close to the GW source, then it could alter the generation of the GW signal itself.[5] In this paper we focus the discussion only on the

[4] In this this work we focus on LISA as the primary space-based GW observatory example, but we also want to acknowledge that other missions such as Taiji or TianQin could enable similar GW observations (Z. Luo et al. 2020; E.-K. Li et al. 2025).

[5] For example, true *triple*-body GW sources may be very important for future consideration, especially in preparation for the LISA mission. And other studies, some predating N. Seto (2008), investigated direct GW detection with LISA from just a *single* star and a tight-orbit exoplanet (see, e.g., V. Ferrari et al. 2000; K. W. K. Wong et al. 2019; A. Kurban et al. 2026). However, these studies suggest that this possibility is less likely.

Doppler modulation scenario and avoid considering the scenario where the orbiting body directly contributes to the generation of the GW signal (which would likely be negligible in many cases). Furthermore, we will exclusively use the term "exoplanet" when describing the perturbing body. However, as we will see, the perturbing body may not be a planet, but could instead be a brown dwarf or stellar companion. And lastly, we will focus on just a single exoplanet. Multi-exoplanet systems would create a more complicated RV observation of the GW source.

With traditional EM observations of stars, the star produces an essentially continuous EM spectrum, with missing absorption lines. If an exoplanet is in orbit around the star and causes the star to move radially along the observer's line of sight, then over time this RV motion creates an observable Doppler modulation of the star's absorption lines. The first exoplanet around a Sun-like star was discovered using this RV method (M. Mayor & D. Queloz 1995). This has elevated EM RV to one of the most prolific methods for discovering and studying exoplanets to date.

With GW observations of stars, if the GW source is nearly monochromatic (i.e., it produces a nearly constant frequency GW), then the source produces a *single* spectral line in the GW spectrum.[6] Therefore, all of the observed signal's power will be found at the GW frequency, which we will refer to as the "carrier." If an exoplanet orbits the GW source, we can think of the exoplanet's orbital frequency as the "message." And if that exoplanet causes the GW source to move radially along the observer's line of sight, then this Doppler modulates the GW frequency. Therefore, while the observed power of the GW signal still peaks at the carrier frequency, the exoplanet's presence attaches additional "messages" at integer multiples of the message frequency in the sidebands *around* the carrier. This phenomenon is well understood in signal analysis and telecommunications.

*The analogy to telecommunications: carriers and messages.* In telecommunications, the goal is to transmit low-frequency "messages" (e.g., speech or data information) that have been converted to low-frequency EM waves. For example, human ears can perceive sound ranging from $\mathcal{O}(10-10,000)$ Hz in the audio spectrum. This range of sound frequencies converted directly into light corresponds to low-frequency radio in the EM spectrum. It is typically impractical, however, to transmit these low-frequency radio messages directly (due to power and cost considerations). However, sending and receiving higher-frequency radio waves is easier and more convenient.

Therefore, to send the low-frequency message, we begin by first generating a high-frequency EM "carrier" signal. This is an EM wave that in and of itself does not contain any important message—it is just a constant-frequency wave. But the carrier is intentionally generated at the higher frequency, where it is easier (practically) to transmit/receive. Next, we modulate that carrier wave by the desired low-frequency message and transmit the final *modulated* signal. In doing this, all of the message information is attached to the carrier and can be found in the sidebands around the carrier frequency. Finally, once a receiver picks up this high-frequency modulated signal, the signal is demodulated, and the low-frequency message is recovered. The received message can then be converted back into the audio spectrum.

Two commonly used techniques for the modulation method itself are amplitude modulation (AM) of the carrier signal (where the signal's amplitude is the part that is modulated) and frequency modulation (FM) of the carrier signal (where the signal's frequency is the part that is modulated)—hence the terminology of "AM" and "FM" radio stations, which is referencing how the carrier signal was first modulated before transmission.[7]

*This study.* The primary goal of this study is to investigate some of the physics principles of the GW RV method itself. Other work, such as M. L. Katz et al. (2022), has developed more complex physical models and performed the first Bayesian analyses to try to search for and extract exoplanet signals from synthetic LISA datasets. And a lot of focus has gone into the development of population synthesis models to help us predict and understand how many potential exoplanets might be present around the specific DWD systems (G. Columba et al. 2023; S. Ledda et al. 2023; A. Nigioni et al. 2026).

But in this work we explore more of the details of the fundamental physics. And while most work so far focuses on DWDs as the candidate system with LISA as the detection observatory, we approach the problem more agnostically, by providing multiple GW source types that GW RV could apply to and by expanding the scope to cases for terrestrial GW observatories.

The structure of this paper is as follows: Section 2 provides the development of the GW RV method and model. This includes a detailed discussion of the fundamental similarities and differences between EM and GW RV and models for two candidate exoplanet host systems (binaries and triaxial bodies). Section 3 then provides a detailed analysis of the effect of the exoplanet on the observed GW signal spectrum. It motivates an exoplanet discovery space by investigating how the power spectrum and signal-to-noise ratio (SNR) change, considers the additional Doppler effects induced by the GW observatory's motion, and investigates the potential for exoplanet-induced bias in the measurements of other system parameters. Lastly, Section 4 summarizes the paper. The appendices contain additional useful information for the curious reader. While Appendix A provides additional contextual information for Section 2, the reader may wish to read Appendices B and C before Section 3 in order to gain insight into how to derive the analytic calculations.

Finally, all figures/calculations presented in this paper can be reproduced using the code developed for this work, which has been made publicly available (see *Materials Availability* in the Acknowledgments).

## 2. The Gravitational-wave Radial Velocity Method

### 2.1. Model Considerations and Assumptions

*Orbital dynamics.* As an exoplanet orbits a GW source, it will cause the center of mass of the system to wobble, thereby producing a Doppler modulation on the observed frequency of the GW. This changing frequency will therefore result in a

[6] These lines will have some finite width to them (since the GW frequency will evolve over time) and other features as we show later in this paper. But generally each one of these sources will make a single "line" in the overall GW spectrum.

[7] AM broadcasting falls around $\mathcal{O}(100-1000)$ kHz, and FM broadcasting falls around $\mathcal{O}(10-100)$ MHz. These are medium and very high frequency radio ranges, respectively. Considering again the example of sending human audible messages, we see that indeed these EM carrier frequencies are many orders of magnitude higher in frequency than the messages attached to them.

Doppler modulation on the observed phase of the GW. For the purposes of this study we assume a circular orbit for the exoplanet.

In principle, the effect of the exoplanet could be more nuanced than the treatment shown here for *binary* GW systems. For example, J. H. Steffen et al. (2018) investigated the effect of tidal modulations on the inner binary caused by an orbiting exoplanet companion. As another example, if the exoplanet has sufficient mass and passes very close to the binary (e.g., if it has a highly eccentric orbit), then the gravitational attraction of the exoplanet could disturb the orbits of the other two, resulting in a more complicated gravitational waveform. In three-body physics, this is the well-studied Kozai–Lidov mechanism (see, e.g., S. Naoz 2016). Therefore, it will be important for future work to incorporate eccentricity and effects like this into the modeling of binary GW sources with potential exoplanets, in order to understand what kind of effect this might have on the observability and detection of these systems (see, e.g., T. Robson et al. 2018; M. L. Katz et al. 2022). Modeling eccentricity and its effect on *triaxial* GW systems will also be important in future work.

Additionally, for the binary systems, we also assume circular orbits for the emitting GW system. GW systems with nonnegligible eccentricity emit GWs at multiple harmonics, which further complicates their waveforms and this analysis.

*System evolution.* GW sources by their nature have frequencies that evolve over time, as the radiation of GWs carries energy out of the system. Colloquially, this is referred to as "chirping." If the GW system is a binary, then this loss of energy causes the binary's orbit to shrink and spin up, and hence the chirp evolves to higher frequencies. If the GW system is an isolated triaxial body, then this loss of energy causes the body to spin down, and hence the chirp evolves to lower frequencies. Chirping results in both phase and amplitude evolution of the GW. However, as long as the system is not near inspiral or merger, phase chirping will dominate. For the purposes of this study, we focus on systems being purely driven by GW radiation and phase chirping (e.g., detached DWDs).[8]

In principle, there could be more complicating physics at play in real systems, such as mass transfer and accretion disks (K. Kremer et al. 2017), pulsar braking (D. R. Lorimer & M. Kramer 2012), or pulsar glitches (D. Antonopoulou et al. 2022), which may result in deviations from the frequency evolution expected under purely GW-driven systems (see additional discussion in Appendix A). Future work will also benefit from considering the effects different environments might have on the observability and detection of these systems.

### 2.2. The Doppler Effect

To begin constructing the GW model, we start with the (orbital) angular frequency of the GW system in its own reference frame. We use a Taylor expansion approximation of the frequency to capture the fundamental monochromatic frequency $\omega_0$ plus the chirp term $\dot{\omega}_0$:

$$\omega(t) \approx \omega_0 + \dot{\omega}_0(t - t_0). \tag{1}$$

Here we assume that the evolution of the system is slow, and hence the chirp is only dominated by the first derivative term, which should suffice for most systems.[9] It is important to note that the development of the Doppler effect on the GW model in this section is mostly agnostic of the type of source that is generating the GW. Section 2.4 is where we introduce the actual GW source: either a binary system or a triaxial body. Until that section, the only dependence of source type in the discussion here is the significance of $\omega$—it is referring either to the orbital angular frequency, if the source is a binary, or to the angular frequency, if the source is a triaxial body.

To the observer of the GW system, the exoplanet's RV $V_r$ (i.e., the component of the exoplanet's velocity along the observer's line of sight) determines the strength of the Doppler modulation of the system's center of mass:

$$V_r(t) \equiv K\cos(\phi_p + 2\pi f_p(t - t_0)), \tag{2}$$

where $\phi_p$ and $2\pi f_p = \omega_p$ are the initial phase and orbital frequency of the exoplanet, respectively. Thus, the Doppler effect tells us that the *observed* frequency relates to the source-frame frequency via

$$\begin{aligned}\omega_{\rm obs}(t) &= \left[1 + \frac{V_r(t)}{c}\right]\omega(t),\\ &\approx \left[1 + \frac{K}{c}\cos(\phi_p + 2\pi f_p(t - t_0))\right]\\ &\quad\times[\omega_0 + \dot{\omega}_0(t - t_0)].\end{aligned} \tag{3}$$

The "semiamplitude" parameter $K$ is the maximum radial speed for the system, and crucially it encodes several key pieces of information about the exoplanet (V. Bozza et al. 2016). Here $c$ is the speed of light. For an exoplanet in a circular orbit about the GW system, this parameter is

$$\begin{aligned}K &= \left(\frac{2\pi G}{P}\right)^{1/3}\frac{M_p\sin(i)}{(M_{\rm sys} + M_p)^{2/3}},\\ &\approx (28.44\ {\rm m\ s^{-1}})\left(\frac{1\ {\rm yr}}{P}\right)^{1/3}\left(\frac{M_p}{M_{\rm Jup}}\right)\\ &\quad\times\left(\frac{M_\odot}{M_{\rm sys} + M_p}\right)^{2/3}\sin(i),\end{aligned} \tag{4}$$

where $P = 1/f_p$ is the exoplanet orbital period, $M_p$ and $M_{\rm sys}$ are the exoplanet and system masses, respectively ($M_{\rm sys} = M_1 + M_2$ for the binary case and $M_{\rm sys} = M_1$ for the triaxial body case), $i$ is the inclination of the exoplanet's orbit to the line perpendicular to the line of sight, and $M_\odot$ and $M_{\rm Jup}$ are the mass of the Sun and Jupiter, respectively. The maximal Doppler effect occurs when the system orbits edge-on to the line of sight (i.e., $i = \pi/2$), and no Doppler effect occurs if the exoplanet orbits perpendicular to the line of sight (i.e., $i = 0$).

[8] Typically in the GW community, the term "chirp" specifically implies the rapid (nonlinear-in-time) evolution of the GW source frequency from a low to a high frequency. In this study, however, our use of the word "chirp" will specifically only include the linear-in-time term (see Equation (1)).

[9] But, for example, a binary system nearing its merger phase should include a more accurate frequency model. Such might be the case of a BNS system with an exoplanet. However, in this scenario, by the time the system nears the inspiral-to-merger phase, the influence of the exoplanet would likely be negligible, as this phase would be much shorter than the exoplanet's orbital period. To improve the chance of detection of the exoplanet via GWs, the chirp should be evolving at a slower rate.

GW measurements and observables map to the phase of the GW itself (see further discussion in Section 2.3). Therefore, we need to use the observed Doppler-shifted frequency to calculate the observed Doppler-shifted phase. Under the assumptions described in Section 2.1, the emitted GW frequency is twice the (orbital) angular frequency of the system (in both binary and triaxial body systems): $f_{\rm GW} = 2f = \omega/\pi$.[10] Therefore, to calculate the observed GW phase, we need to integrate the observed GW frequency:

$$\Phi_{\rm GW,obs}(t) = \int \frac{d\Phi_{\rm GW,obs}(t)}{dt} dt = \int 2\omega_{\rm obs}(t) dt. \tag{5}$$

Plugging in Equation (3) and integrating by parts gives the result

$$\begin{aligned}\Phi_{\rm GW,obs}(t) &= \Phi^{\rm Mono}_{\rm GW,obs}(t) + \Phi^{\rm Chirp}_{\rm GW,obs}(t),\\ &\approx \overbrace{\phi_0 + 2\omega_0(t-t_0) + \underbrace{\epsilon\Big[\sin(\phi_p + \omega_p(t-t_0)) - \sin(\phi_p)\Big]}_{\text{Exoplanet Doppler Effect}}}^{\text{Monochromatic GW}}\\ &\quad \underbrace{+\, \dot{\omega}_0(t-t_0)^2 + \epsilon_{\rm chirp}\overbrace{\Big[\omega_p(t-t_0)\sin(\phi_p+\omega_p(t-t_0)) + \cos(\phi_p+\omega_p(t-t_0)) - \cos(\phi_p)\Big]}^{\text{Exoplanet Doppler Effect}}}_{\text{Chirp}}\\ &\approx \overbrace{\phi_0 + 2\omega_0(t-t_0) + \underbrace{\epsilon\Big[\sin(\phi_p+\omega_p(t-t_0)) - \sin(\phi_p)\Big]}_{\text{Exoplanet Doppler Effect}}}^{\text{Monochromatic GW}} + \underbrace{\dot{\omega}_0(t-t_0)^2}_{\text{Chirp}}.\end{aligned} \tag{6}$$

Writing it in this way, we can see how all of the components/assumptions separate out and how they will contribute to the overall phase. The phase change of the GW by the exoplanet is controlled by the "phase modulation index" (PMI) parameter:[11]

$$\begin{aligned}\epsilon &\equiv \frac{K}{c}\frac{2\omega_0}{\omega_p} = \frac{K}{c}\frac{f_{\rm GW,0}}{f_p} = \frac{(2\pi G)^{1/3}}{c}P^{2/3}\\ &\quad \times \frac{M_p \sin(i)}{(M_{\rm sys}+M_p)^{2/3}} f_{\rm GW,0},\\ &\approx (2.992\ {\rm rad})\left(\frac{P}{1\ {\rm yr}}\right)^{2/3}\left(\frac{M_p}{M_{\rm Jup}}\right)\\ &\quad \times \left(\frac{M_\odot}{M_{\rm sys}+M_p}\right)^{2/3}\left(\frac{f_{\rm GW,0}}{1\ {\rm Hz}}\right)\sin(i).\end{aligned} \tag{7}$$

The PMI is the critical parameter in our analysis, as it is ultimately the parameter that controls the strength of the exoplanet Doppler effect on our signal. This parameter offers a great deal of insight into the fundamental physics of the GW RV method (discussed further in Section 2.3).

[10] Throughout this paper we will reserve $\omega$ to denote the astrophysical system's (orbital) angular frequency (both source and exoplanet). When discussing the GW frequency itself, we will typically denote it by $f_{\rm GW}$.

[11] This terminology appears in M. Maggiore (2007) when discussing the Doppler effect, which in general is a helpful reference to this work.

The "chirp modulation index" parameter $\epsilon_{\rm chirp}$ takes a very similar form:

$$\begin{aligned}\epsilon_{\rm chirp} &\equiv \frac{K}{c}\frac{2\dot{\omega}_0}{\omega_p^2} = \frac{K}{c}\frac{\dot{f}_{\rm GW,0}}{2\pi f_p^2} = \frac{(2\pi G)^{1/3}}{2\pi c}P^{5/3}\\ &\quad \times \frac{M_p \sin(i)}{(M_{\rm sys}+M_p)^{2/3}}\dot{f}_{\rm GW,0}\\ &\approx (4.76\times 10^{-10}\ {\rm rad})\left(\frac{P}{1\ {\rm yr}}\right)^{5/3}\left(\frac{M_p}{M_{\rm Jup}}\right)\\ &\quad \times \left(\frac{M_\odot}{M_{\rm sys}+M_p}\right)^{2/3}\left(\frac{\dot{f}_{\rm GW,0}}{1\ {\rm nHz\ yr^{-1}}}\right)\sin(i).\end{aligned} \tag{8}$$

However, the approximation given by the final line of Equation (6) holds when

$$\omega_0 \gg \frac{\dot{\omega}_0}{\omega_p} \xrightarrow{\text{i.e.}} f_{\rm GW,0} \gg \frac{\dot{f}_{\rm GW,0}}{2\pi f_p} = \frac{\dot{f}_{\rm GW,0}}{2\pi}P. \tag{9}$$

Under this condition, we can neglect any FM terms that appear in the chirp phase. Therefore, we are going to ignore the contribution of the chirp modulation index in our analysis here since it controls a subdominant effect.

It is in this final form of Equation (6) that we now have just a general phase chirping, FM signal (see Appendix C). The carrier frequency is the GW frequency: $\omega_c = 2\omega_0 = 2\pi f_{\rm GW,0}$. The message frequency is the exoplanet's orbital frequency: $\omega_m = \omega_p = 2\pi f_p$. Finally, just to reemphasize, to apply the results derived in Appendix C, in this work we are assuming that the GW carrier frequency is higher than the exoplanet message frequency, $f_{\rm GW,0} > f_p$ (which matches our earlier telecommunications analogy).

*Caveat: the infinite exoplanet period limit.* In Equation (7) (and Equation (8)) it appears that in the limit that the exoplanet's period $P \to \infty$, the PMI $\epsilon \to \infty$. Conceptually this should not be the case, as the infinite exoplanet period limit should be equivalent to a system without an exoplanet. We also see that the condition for when we can neglect the chirp modulation index terms (Equation (9)) no longer holds.

Therefore, in this particular limit we need to treat the analysis of our expression in Equation (6) more carefully. If we Taylor-expand the collective "Exoplanet Doppler Effect" terms in the second line of Equation (6) in terms of the dimensionless parameter $\omega_p(t - t_0) = 2\pi(t - t_0)/P$, then we find that indeed in the limit that $t/P \rightarrow 0$ the exoplanet effect vanishes as expected.

This analysis emphasizes another key insight. Equation (7) shows us that the PMI increases with longer orbital period exoplanets. However, the sinusoidal terms in the square brackets will suppress this greater PMI if the orbital period $P$ is significantly longer than the observation time. This is a familiar concept for periodically varying signals, namely that to observe a periodic change in a signal, the observation time should likely be several times longer than the signal's period.

### 2.3. Radial Velocity: EM versus GW

The above relationships reveal an intriguing and key difference between EM-based RV observations and GW-based RV observations. In EM RV, we are able to directly relate the observable to the semiamplitude $K$ of the exoplanet. From the Doppler effect (first line of Equation (3), rewritten in terms of the wavelength $\lambda$), we have

$$V_r(t) = \frac{c(\lambda(t) - \lambda_{\rm obs}(t))}{\lambda_{\rm obs}(t)}. \tag{10}$$

Thanks to the very nature of the EM-based observation (spectral "fingerprints" from known atomic elements), we can match the star's observed absorption lines to lab-based measurements of the same absorption features. This allows us to match model templates of Equation (2) and directly estimate the parameters found within $K$. From the form of Equation (4), we see that EM RV is therefore sensitive to high-mass, short-period exoplanets.

However, in GW RV we fundamentally lack lab measurements of the observed spectral features.[12] Therefore, GW RV is not directly sensitive to the semiamplitude parameter $K$. Instead, in GW analysis, we match the laser interferometer observables directly to model templates that contain the GW phase given in Equation (6). And crucially, the phase is an *integrated* effect—i.e., the Doppler effect caused by the exoplanet gets added up and integrated over time. GW RV is therefore sensitive to the PMI $\epsilon$. It is thanks to the time integration that this parameter contains an *additional* factor of $P$, as compared to the semiamplitude. And this extra $P$ is enough to flip the proportionality. From the form of Equation (7), we see that GW RV is therefore sensitive to high-mass, *long-period* exoplanets.[13] This behavior can be seen later in Figures 3 and 5.

So conceptually, while exoplanets close to the EM or GW source will produce a larger wobble (larger $K$) thanks to the stronger gravitational interaction, the buildup over time of the excess phase is small because the exoplanet is orbiting rapidly. This is good for EM RV because we can directly model $K$. At greater orbital separations, the wobble is smaller (weaker gravitational interaction), but since the orbit takes longer, there is more time for the phase deviation to build up (larger $\epsilon$). And this is better for GW RV because we can directly model $\epsilon$.

[12] The observed GW spectrum is being generated by stars after all, not atoms.

[13] Of course, as mentioned in the caveat in Section 2.2, the observation time of the experiment should still be several times longer than the underlying exoplanet period in order to enable a statistically significant detection.

### 2.4. The GW Models for Binary Systems and Triaxial Bodies

GWs cause a time-varying strain $h$ on the surrounding spacetime as they propagate, analogous to the oscillating electric and magnetic fields of an EM wave. The resulting strain on spacetime results in alterations to the path of propagation of photons, an effect that can be exploited and observed. This is why GW observatories are typically designed as sophisticated interferometers.

Mathematically, the strain is represented by the GW waveform. General relativity predicts the two GW polarizations that have been observed and measured to date, the "plus" and "cross" polarizations (denoted by subscript $+$ and $\times$). While alternative theories of gravity predict that there may exist other modes (see, e.g., C. M. Will 2014), we focus here only on these two. The orientation of the GW system with respect to the observer's line of sight can result in the observer measuring a mixture of these two polarizations (analogous to an elliptical polarization in EM, and denoted here with an $H$). The GW polarizations can be expressed in the following manner:

$$\begin{aligned} h_+(t) &= -h\frac{(1+\cos^2(\iota))}{2}\cos(\Phi_{\rm GW,obs}(t)),\\ h_\times(t) &= -h\cos(\iota)\sin(\Phi_{\rm GW,obs}(t)),\\ H_+(t) &= h_+\cos(2\psi) - h_\times\sin(2\psi),\\ H_\times(t) &= h_+\sin(2\psi) + h_\times\cos(2\psi), \end{aligned} \tag{11}$$

where the strain amplitude $h$ is determined by which type of system we are observing (binary or triaxial body; Equations (12) and (14)), $\iota$ is the system's inclination angle, and $\psi$ is the system's polarization angle (both $\iota$ and $\psi$ are Euler angles that describe the orientation of the GW system to the observer). Therefore, $\iota = 0$ represents a system the observer is viewing face-on, and $\iota = \pi/2$ represents a system the observer is viewing edge-on.[14] Equation (C12) in Appendix C.5 is the fully analytic *frequency domain* version of Equation (11), which includes the exoplanet's Doppler effect. This could be very useful for future modeling work.

In general, for the two sources of GWs that we are considering in this work—binary systems and triaxial bodies—the only aspect of the model that changes between them is the strain amplitude $h$.

*Binary system.* Binary systems of ultracompact objects—white dwarfs, neutron stars, and black holes—are excellent sources of GWs (G. Nelemans et al. 2009). Their small size allows them to exist in orbits that are so tight that they enter the regime where they can radiate energy out of the system in the form of GWs, with orbital periods <1 hr.[15]

[14] Note that, given the way we have defined the model in this work, the inclination angle of the exoplanet $i$ is not coupled to the inclination angle of the GW system $\iota$. These parameters are kept independent here. So, for example, if the exoplanet orbits in the same plane as the binary and the binary is seen edge-on, then both $\iota = i = \pi/2$.

[15] Really any massive binary system where the objects are compact enough that they can orbit each other with a period of <1 hr could fall into this category. This might include objects such as low-mass main-sequence stars, or the helium cores of giant stars that have lost their hydrogen envelope (Y. Götberg et al. 2020; M. Colpi et al. 2024).

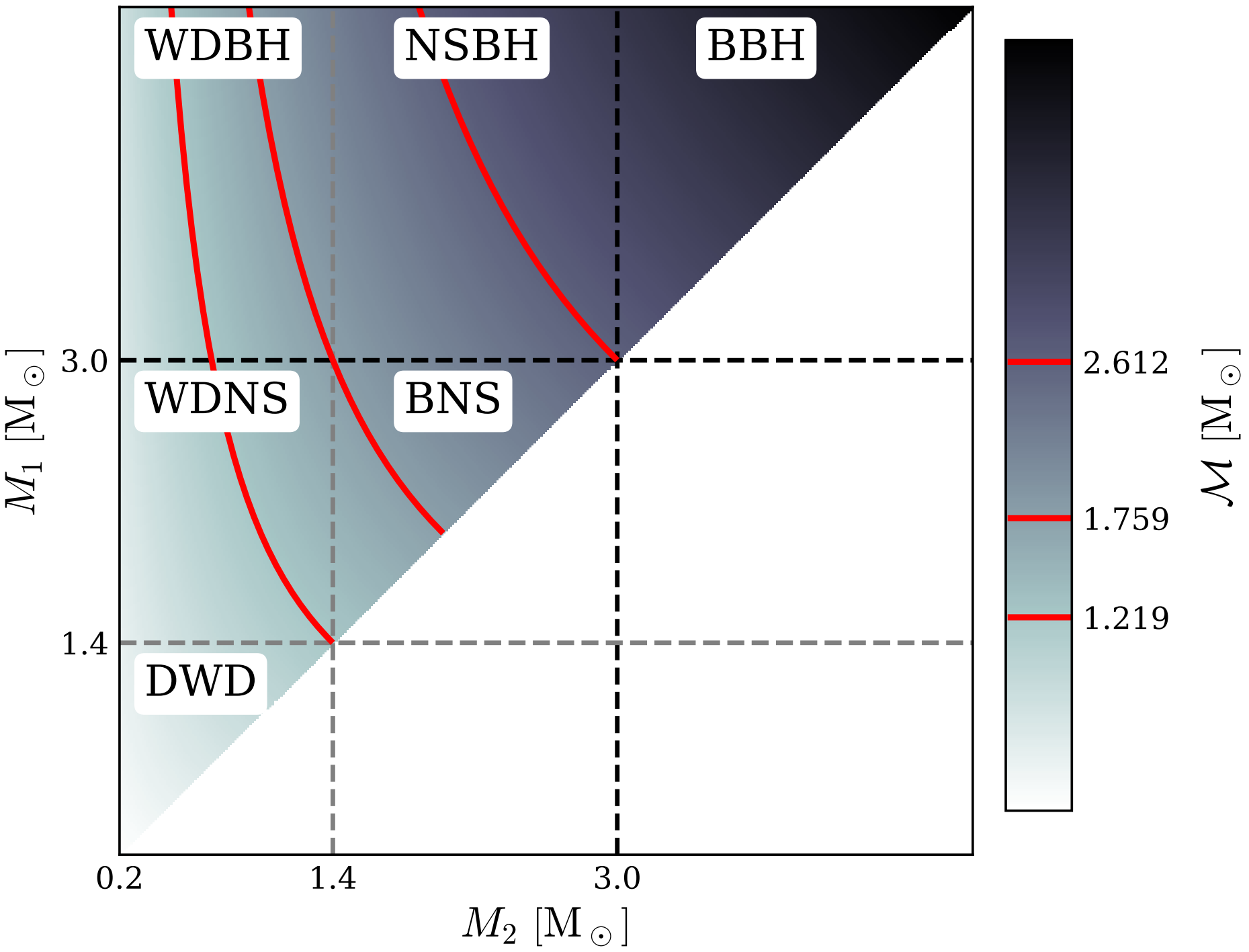

**Figure 1.** Ultracompact binaries for GW RV discovery space. GW measurements are sensitive to the chirp mass $\mathcal{M}$, a degenerate combination of the individual star masses $M_1$ and $M_2$. Here it is assumed that $M_1 \geqslant M_2$ (the plot is otherwise symmetric about the diagonal). The boundaries between white dwarfs, neutron stars, and black holes have been informed by both theory and observation. While these are not hard boundaries, we use the fiducial values of 0.2–1.4 $M_\odot$ for the regime of white dwarfs, 1.4–3 $M_\odot$ for the regime of neutron stars, and $\geqslant$3 $M_\odot$ for the regime of black holes. Therefore, a GW measurement of $\mathcal{M} \lesssim 1.219\ M_\odot$ would suggest that at least one star is a white dwarf; $1.219\ M_\odot \lesssim \mathcal{M} \lesssim 1.759\ M_\odot$ would suggest that the system is a WDNS, WDBH, or BNS; $1.759\ M_\odot \lesssim \mathcal{M} \lesssim 2.612\ M_\odot$ would suggest that the system is a WDBH, BNS, or NSBH; and $\mathcal{M} \gtrsim 2.612\ M_\odot$ would suggest that at least one star is a black hole (see text for the explanation of the abbreviations). Of these systems, the LISA mission anticipates being able to individually resolve $\mathcal{O}(10^4)$ DWDs throughout the Milky Way. If the other systems with greater chirp mass exist in the Milky Way, they will likely produce loud signals in LISA.

The strain amplitude of a binary system can be expressed in the form

$$h_{\text{binary}} = 2\underbrace{\left(\frac{2(G\mathcal{M})^{5/3}\omega_0^{2/3}}{c^4 D}\right)}_{\equiv A_{\text{binary}}} \times \left[1 + \frac{\dot{\omega}_0}{\omega_0}(t - t_0)\right]^{2/3} \approx 2A_{\text{binary}}. \quad (12)$$

In this study we focus on phase chirping systems only. We will neglect the $\mathcal{O}\left(\frac{\dot{\omega}_0}{\omega_0}\right)$ term in Equation (12), which represents the secular change in the GW amplitude due to the chirp. As long as the system is not near the inspiral-to-merger phase, the dominant effect of the chirp is contained in the GW phase only (Equation (6)). The distance to the source is $D$, and the chirp mass of the binary system is defined as

$$\mathcal{M} \equiv \frac{(M_1 M_2)^{3/5}}{(M_1 + M_2)^{1/5}}, \quad (13)$$

a degenerate combination of the individual masses $M_1$ and $M_2$ of the binary.

GW RV could enable a new class of exoplanet discovery—post-main-sequence circumbinary systems. There are six potential systems in this discovery space: DWDs, binary neutron stars (BNSs), binary black holes (BBHs), white dwarf–neutron stars (WDNSs), white dwarf–black holes (WDBHs), and neutron star–black holes (NSBHs).[16] This system space as a function of the GW-measured chirp mass is shown in Figure 1.

At these subhour orbital periods, these systems would be likely candidates for mid-frequency GW observatories, in the mHz part of the GW spectrum. Of these six candidate systems, ultracompact binaries primarily consisting of DWDs in the Milky Way are expected to be the single most abundant type of object that will be discovered by the upcoming LISA mission (M. Colpi et al. 2024). It is estimated that $\mathcal{O}(10^7)$ of these systems exist in the Milky Way in orbits that would generate GWs in the part of the spectrum LISA will be sensitive to and that $\mathcal{O}(10^4)$ of these systems will be individually resolved and classified by LISA (K. Breivik et al. 2020).[17] Therefore, it is of keen interest to understand whether exoplanet signals from these sources could be measured. And although it is estimated that there will be many fewer systems with neutron stars or black holes (possibly $\mathcal{O}(10 - 100)$), if they do indeed exist, they will likely produce relatively loud signals (since their chirp masses are larger, and hence the strain amplitude $h_{\text{binary}}$ will be larger).

The colloquial name "Magrathea" worlds has been given to these DWD circumbinary systems (C. Danielski et al. 2019; N. Tamanini & C. Danielski 2019; G. Columba et al. 2023;

[16] These are some of the notations used in GW literature, but they are not exclusive. As just one example, sometimes "BHB" is used for "black hole binary."

[17] Most of these systems will contribute to a "Galactic foreground noise" for the mission and will not be distinguishable from one another.

S. Ledda et al. 2023; A. Nigioni et al. 2026).[18] And while an exoplanet in such a DWD system has yet to be discovered, there certainly is interesting precedent for their possible existence. Numerous circumbinary exoplanets have been confirmed to date (see, e.g., P. Thebault & D. Bonanni 2025). And several exoplanets and exoplanet candidates around isolated white dwarfs have been discovered (B. T. Gänsicke et al. 2019; A. Vanderburg et al. 2020; J. W. Blackman et al. 2021; M. A. Limbach et al. 2024, 2025; S. E. Mullally et al. 2024).

*Triaxial body.* Isolated single-body ultracompact objects may also be able to generate GWs, but to do so, they must have some asymmetry in their physical shape. Therefore, these objects are typically modeled as triaxial ellipsoids. The current strongest candidate for this type of GW source are neutron stars—often colloquially referred to as "neutron stars with mountains." As a simplistic model, here we will assume that the rotational axis aligns with one of the principle axes of the ellipsoid (chosen here to be labeled the $z$-axis), to avoid the effects of precession on the GW waveform.

The strain amplitude of a triaxial body can be expressed in the form

$$h_{\rm triax} = 2\underbrace{\left(\frac{2GeI_z\omega_0^2}{c^4D}\right)}_{\equiv A_{\rm triax}}\left[1 + \frac{\dot{\omega}_0}{\omega_0}(t - t_0)\right]^2 \approx 2A_{\rm triax}. \quad (14)$$

As with the binary case, in this study we focus on phase chirping systems only and will neglect the $\mathcal{O}\left(\frac{\dot{\omega}_0}{\omega_0}\right)$ term in the amplitude. The system's principle moment of inertia (about the $z$-axis) is $I_z$, and its ellipticity parameter is defined as $e \equiv (I_x - I_y)/I_z$.

From our observations of pulsars, we know that neutron stars tend to rotate very rapidly, down to millisecond periods. At these spin rates, this would make them likely candidates for high-frequency GW observatories, in the kHz part of the GW spectrum. This is where the current LIGO–Virgo–KAGRA (LVK) GW observatories operate. To date, no such GW signals have been detected by these observatories. Both targeted searches of known pulsars and untargeted searches across the sky for potential new neutron stars have yet to reveal a GW signal of this kind (see, e.g., R. Abbott et al. 2022; A. G. Abac et al. 2025a). Although these nondetections are placing stricter limits on the theoretical ellipticity of triaxial bodies, future GW detectors such as the Einstein Telescope and Cosmic Explorer observatory that would far surpass the current design sensitivity of the LVK are being envisioned.

Interestingly, we have detected a small number of confirmed exoplanets around pulsars thanks to pulsar timing methods.[19] But also pulsar timing has discovered hundreds of companion *objects* orbiting in pulsar systems (see, e.g., the Australia Telescope National Facility (ATNF) Catalogue v2.6.5, and P. F. Wang et al. 2025). So there exists precedent for the possibility of exoplanets in triaxial neutron star systems, if such neutron stars themselves exist.

*GW source distance.* It is important to point out in Equations (12) and (14) how the amplitude of the GW is inversely proportional to the source distance $1/D$. This is a critical distinction compared to EM waves, whose amplitudes are inversely proportional to the source distance squared $1/D^2$. The significance of this is that although GWs are generally much weaker and harder to detect, they will have greater reach through the Galaxy and the Universe.[20] This gives intriguing motivation for why it is important to think about and search for exoplanets orbiting GW sources, because in principle we may be able to detect them at distances much farther away from Earth as compared to our current ability to detect them through EM channels alone (C. Danielski & N. Tamanini 2020).

### 2.5. GW RV Enabled Exoplanet Measurements

In addition to the model parameters that have been shown so far, each system will have two sky angles $\beta$ and $\lambda$. The GW strain does not depend on the position on the sky, but rather on the GW *response* of the detector. To understand how sensitive a GW detector is to a specific source, one needs to use the strain to write out a model for the response. In this study we are primarily focusing on an analysis that is agnostic of a specific detector. But Section 3.3 does show some simulations specific to the LISA design.

In summary, a GW RV observation of one of these exoplanetary systems would allow us to make measurements of the following parameters:[21]

$$\begin{aligned} &\text{GW Source:} \{A,\ f_{\rm GW,0},\ \dot{f}_{\rm GW,0},\ \iota,\ \psi,\ \phi_0,\ \beta,\ \lambda\},\\ &\quad\text{Exoplanet:} \{\epsilon,\ P,\ \phi_p\},\\ &\quad\text{Derived / Constrained:} \left\{\frac{M_p\sin(i)}{(M_{\rm sys}+M_p)^{2/3}}\right\}\\ &\quad+ \underbrace{\{\mathcal{M},\ D\}}_{\text{binary system}} \text{ or } \underbrace{\{e^2I_3,\ eR\}}_{\text{triaxial body}}. \end{aligned}$$

Appendix A shows how the derived/constrained parameters can be obtained from combinations of the measured GW source parameters. And thanks to Equation (7), we note that successful measurements of the GW frequency $f_{\rm GW,0}$, exoplanet period $P$, and PMI $\epsilon$ would give us a measurement of the above exoplanet mass, system mass, and inclination coupled quantity.

Furthermore, the GW observation would give us the inclination of the GW source itself, $\iota$. If it can be reasonably assumed that the exoplanet's orbital plane aligns with the inclination of the GW system, then the GW measurement could break the $\sin(i)$ degeneracy in the mass ratio.

Joint EM observations of the same system could make these systems uniquely multimessenger exoplanets (e.g., see C. McGrath et al. 2026) and potentially provide new information that could strengthen the certainty of the GW observations and/or break further degeneracies. N. Tamanini & C. Danielski (2019) provide an overview of potential synergies with counterpart EM detection methods.

[18] The "Magrathea" colloquialism (reference from *The Hitchhiker's Guide to the Galaxy*) mirrors the "Tatooine" colloquialism (reference from *Star Wars*) in the exoplanet community. The key distinction is that Tatooine worlds describe exoplanets in circumbinary systems of "living" main-sequence stars, while Magrathea worlds would describe exoplanets in circumbinary systems of "dead" post-main-sequence stars.

[19] Famously, the first exoplanet discovery was around a pulsar (A. Wolszczan & D. A. Frail 1992).

[20] Indeed, as of version 4.0 of the "Gravitational-Wave Transient Catalog" (GWTC-4.0; A. G. Abac et al. 2025b), all GW detections to date have been made of systems located *outside* of the Milky Way.

[21] We note that N. Tamanini & C. Danielski (2019) list $K$ (not $\epsilon$) as the GW-enabled RV measured parameter. Indeed, the difference between these two parameters (see Equation (7)) only depends on factors of the GW and exoplanet frequencies, but we argue here in Section 2.3 that the PMI is the more appropriately sensitive parameter for GW RV, as found from the model Equation (6).

Finally, we emphasize that the list of parameters indicated here comes from the model we have shown under the described simplifying assumptions (primarily, circular orbits). More complex models under different assumptions may list different and additional parameters (see, e.g., T. Robson et al. 2018; M. L. Katz et al. 2022).

## 3. Exoplanet Signal Analysis

The upcoming LISA mission will target the "mHz" regime of the GW spectrum, with its greatest sensitivity falling around 1–10 mHz. The current terrestrial LVK GW observatories are sensitive to the "kHz" end of the GW spectrum, with their greatest sensitivity around 100–1000 Hz. Therefore, in the following sections we choose representative GW carrier frequencies at 5 mHz and 100 Hz for the binary systems and the triaxial systems, respectively.

Additionally, the nominal mission lifetime of LISA will be approximately 4 yr, with a possible extension up to 10 yr (M. Colpi et al. 2024). The LVK observatories have operated in periodic observing runs (e.g., the third observing run operated for a roughly 10-month observation) that have now spanned a decade (A. G. Abac et al. 2025b). In the following sections we chose to use 10 and 4 yr as the representative observation times of the simulated spectra in the binary and triaxial bodies, respectively. Note that, in principle, ground-based detectors are capable of accumulating even longer observing times than space-based missions, thanks to the ability to maintain and upgrade the facilities.

In GW observations, the frequency resolution is set by the observation time: $\Delta f = 1/T_{\rm obs}$. The duration of the observation is critical in providing enough frequency resolution to resolve the features in the spectrum caused by both the presence of the exoplanet and the frequency chirping.

### 3.1. Signal Power

To understand the observed GW signal, here we consider the predicted GW spectrum. In this analysis, the power spectral density (PSD) content of the signal reveals the amount of power per frequency contained in the observed spectrum. In order to remain agnostic of an actual GW experiment being used to detect the signal, in this section we simply present a look at the strain PSD and the underlying physics therein. The reader may find it helpful to refer to Appendices B and C before reading the following sections, in order to understand the fundamentals of the terminology and spectral features (refer to Figures C1 and C2 for useful visual breakdowns of many concepts).

Figure 2 presents side-by-side examples of various cases that demonstrate a number of important physics considerations, representing both the binary systems and triaxial bodies. Each panel is highly zoomed in on the part of the GW spectrum where the signal is present. In all cases the monochromatic cases represent idealized signals, for reference and comparison. In reality, we expect chirping in all physical systems (see Section 2.1).

Here we summarize the following key points:

1. Focusing first on the monochromatic signals, the general structure in all cases is the familiar power spectrum of an FM signal. With the GW carrier frequency $f_{\rm GW,0}$ at the center of the spectrum, the exoplanet's message frequency $f_p$ appears in the sidebands at $n$ integer multiples: $f_{\rm GW,0} + nf_p$.[22]
2. With multiple years of observation, we can achieve very high frequency resolution $\Delta f = 1/T_{\rm obs}$ in the GW observations (as shown in each of these panels and later in Figure 4). This can help us distinguish the message harmonics surrounding the GW carrier frequency. In practice, beyond the scope of this study, there may be additional identification challenges if GW signals from multiple sources are close enough to each other in frequency space that they overlap and create "inter-signal" confusion (see, e.g., Figure C2).
3. The power carried at the GW carrier frequency is typically the greatest, except in more extreme cases where the PMI becomes very large, as seen in the top right panel. Here we can see that individual messages in the sideband can end up carrying more power than the GW carrier frequency itself.
4. With the presence of chirping, the spectrum spreads out and grows wider with increasing observation time. For a slowly chirping GW source, that spectrum spans the range $f_{\rm GW,0} + \Delta f_{\rm GW}$, where the total width of the frequency spectrum of the GW chirp over some observation window is just $\Delta f_{\rm GW} = \dot{f}_{\rm GW,0} T_{\rm obs}$. For these purely GW-driven systems, the binaries evolve toward higher frequencies (Equation (A1)), while the triaxial bodies evolve toward lower frequencies (Equation (A2)).
5. Each message in the sideband grows over time at the same rate as the carrier frequency. Normally under purely GW-driven frequency evolution, the rate of frequency evolution is itself frequency dependent (see Equation (A1)), meaning that a GW source with a different initial $f_{\rm GW,0}$ will evolve at a different rate. But all of the sidebands, even though they sit at technically different frequency values in the spectrum, evolve together at the same rate. Possibly this could help differentiate sidebands caused by an exoplanet from different GW sources that happen to sit close to each other in frequency space.
6. If the observation window is long enough that the chirp spectrum grows $|\Delta f_{\rm GW}| \geqslant f_p$, then the messages begin to overlap with one another, as well as with the carrier, creating an "intrasignal confusion" in the observed spectrum. This makes it difficult to distinguish the presence of the message frequency. In Figure 2, the examples in the top row have differentiable sidebands even in the chirping case, but the bottom row has intrasignal confusion. This suggests that exoplanets with a shorter period (and hence higher $f_p$) will be less likely to have this intrasignal confusion.

[22] Considering the telecommunications analogy again, this is another useful aspect of FM spectra. The carrier and message frequencies themselves may be many orders of magnitude different from each other, as they are here in this work. But the actual spectrum of the FM signal spans a narrow range of frequencies surrounding just the carrier frequency. Therefore, we do not have to search an enormous range of frequencies to find the message frequency; we primarily just have to know and identify the carrier frequency, and the message will be found in its immediate sidebands.

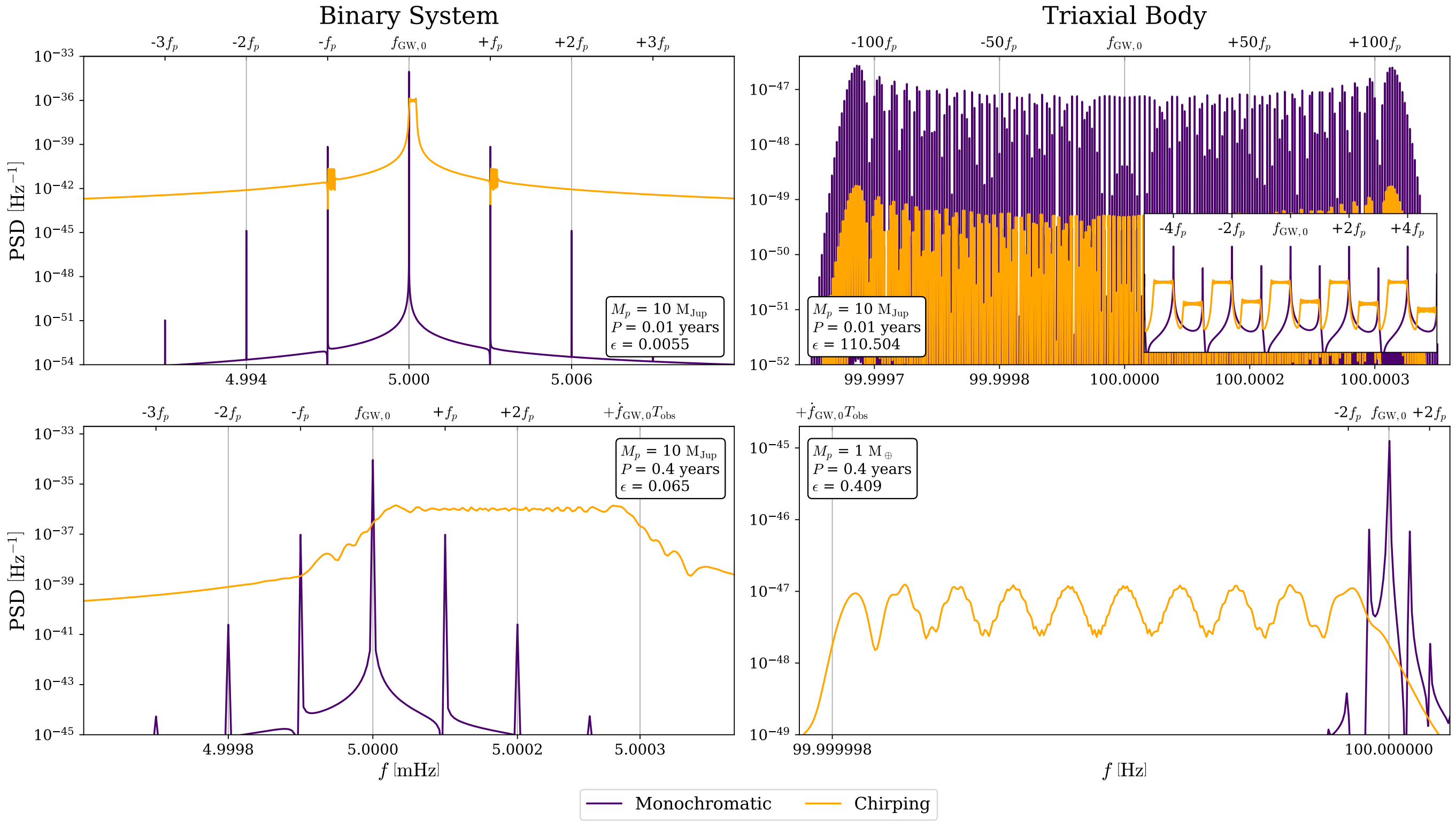

**Figure 2.** The GW spectrum of sources with exoplanets. This is a side-by-side comparison of four simulations, demonstrating the effect of the exoplanet on a GW signal. Specifically these plots are of the PSD of the strain $H_+(t)$. The monochromatic signals represent idealized signals (for comparison)—real GW signals are expected to chirp over time. The chirping simulated here is purely GW driven. For all simulations presented, $D = 2$ kpc, $i = \pi/2$ rad (maximal Doppler effect), and $\iota = \psi = \phi_0 = \phi_{0,p} = 0$ rad. The spectral frequency resolution in each plot is $\Delta f = 1/T_{\rm obs}$ (for the indicated observation times). Note also that all systems here have the same total mass for the GW system, $M_{\rm sys} = 1.4\,M_\odot$ (for the binaries, they are equal in mass). Left column: the binaries here are simulated at $f_{\rm GW,0} = 5$ mHz in the mid-frequency GW spectrum, where the LISA mission will be most sensitive. These simulations are based on an observation time $T_{\rm obs} = 10$ yr. The initial chirp rate is $\dot{f}_{\rm GW,0} = 29.28$ nHz yr$^{-1}$. Right column: the triaxial bodies are simulated at $f_{\rm GW,0} = 100$ Hz in the high-frequency GW spectrum, where terrestrial GW observatories currently operate. These simulations are based on an observation time $T_{\rm obs} = 4$ yr. For the triaxial body simulations, $e = 10^{-6}$ and $I_z = 10^{38}$ kg m$^2$. The initial chirp rate is $\dot{f}_{\rm GW,0} = -542.23$ nHz yr$^{-1}$. Top row: the same exoplanet ($M_p$ and $P$) is simulated in both panels, but the PMI changes by a factor of $2 \times 10^4$ because the GW carrier frequency changes by that factor (a result of Equation (7)). Because $|\Delta f_{\rm GW}| < f_p$, there is no intrasignal confusion in the chirping signal. Bottom row: going from the top to the bottom row, we increase the exoplanet period. As with the top row, the PMI increases by $2 \times 10^4$ from the left to the right column, due to the change in the GW carrier frequency, so we also decrease the mass of the exoplanet in the bottom right panel simply to demonstrate an example where we still get similar features to the bottom left panel, but for a much smaller exoplanet (now an Earth-mass planet). Because $|\Delta f_{\rm GW}| > f_p$, now there is intrasignal confusion in the chirping signal.

7. Finally, because the PMI is directly proportional to the GW carrier frequency (see Equation (7)), moving the same exoplanet system from the mHz to the kHz GW spectrum can result in a $\sim\mathcal{O}(10^4 - 10^6)$ increase to the PMI. This has the potential to create a drastic change in the observed spectrum, as seen in the top panel going from the left to the right column. Decreasing the exoplanet mass and/or period lowers the PMI (again thanks to Equation (7)). In the bottom right panel, we see that a much smaller mass exoplanet in the kHz GW regime can still have a higher PMI than an exoplanet with the same period but much higher mass in the mHz GW regime.

### *3.2. The Exoplanet Power Siphon*

An interesting question to ask is, how much signal power is located at all of the peaks in the observed spectrum? We can calculate this analytically using the techniques developed/explained in Appendices B and C:

$$S^{(1)}_{H_a}\left(f = f_{\rm GW,0} + nf_p + \frac{1}{2}\dot{f}_{\rm GW,0}T_{\rm obs}\right) \approx \frac{2A^2 G_a J_n^2(\epsilon)}{|\dot{f}_{\rm GW,0}|\ T_{\rm obs}},$$
$$\text{where } a \in \{+, \times\},\ \ n \in \mathbb{Z}, \tag{15}$$

and
$$\begin{cases} G_+ \equiv \dfrac{1}{4}(1 + \cos^2(\iota))^2 \cos^2(2\psi) + \cos^2(\iota)\sin^2(2\psi), \\ G_\times \equiv \dfrac{1}{4}(1 + \cos^2(\iota))^2 \sin^2(2\psi) + \cos^2(\iota)\cos^2(2\psi), \end{cases} \tag{16}$$

where $A$ is either the amplitude of the binary (Equation (12)) or the amplitude of the triaxial body (Equation (14)).

Therefore, this equation lets us calculate what we would expect the "averaged" peak values[23] to be for the carrier frequency ($n = 0$) and each sideband message. This immediately shows us how the exoplanet alters the signal. Namely, the strength of each component of the signal (the carrier and the message sidebands) is modified by a Bessel function $J_n(\epsilon)$, which depends exclusively on the PMI. And as a sanity check, in the limit that the exoplanet vanishes (i.e., $\epsilon \to 0$), all of the message sidebands disappear, leaving only the carrier signal, just as expected.[24]

[23] Which we take to be halfway through the GW chirp spectrum. Refer to the discussion in Appendix C.2.
[24] See Appendix C.4.

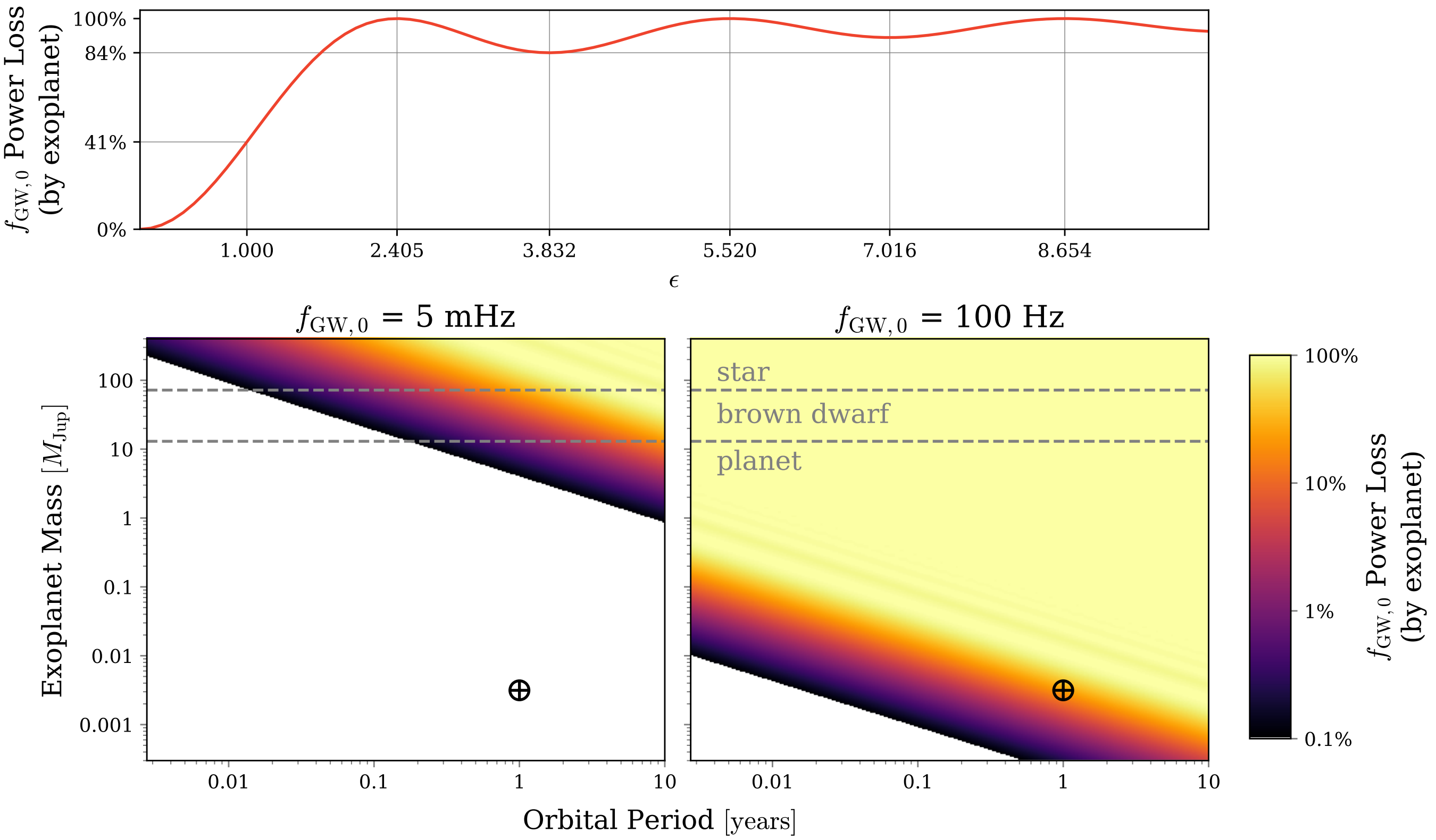

**Figure 3.** An exoplanet discovery space proxy—power siphoned from the GW carrier frequency. The presence of an exoplanet siphons power out of the GW carrier frequency and distributes it into the sideband exoplanet messages. The relative difference in the GW carrier frequency power is given by Equation (17) and plotted here. Top: the loss of power at the carrier frequency is small for $\epsilon \ll 1$. In order to observe significant power loss at the carrier caused by an exoplanet, $\epsilon \gtrsim \mathcal{O}(1)$. Bottom: here we fix the GW carrier frequency and the system mass and show the discovery space as a function of exoplanet mass and period. In both cases, $M_{\rm sys} = 1.5\,M_\odot$ (a value chosen because it makes sense either as an individual neutron star or as the combined mass of several types of the binary systems shown in Figure 1), and the exoplanet's inclination was set to $i = \pi/2$ (maximal Doppler effect). We chose to only show contours down to a relative difference of 0.1%. Dashed lines indicate the boundaries between planets ($M_p \lesssim 13 M_{\rm Jup}$), brown dwarfs ($13 M_{\rm Jup} \lesssim M_p \lesssim 72 M_{\rm Jup}$), and stars ($M_p \gtrsim 72 M_{\rm Jup}$). Earth is marked for reference. Bottom left: a representation of the mHz GW regime, where the LISA mission will be most sensitive. The loss of power at the GW carrier is generally small and suggests that if the Doppler effect is measured it may be for systems where the "exoplanet" is a brown dwarf or even a small star. Bottom right: a representation of the kHz GW regime, where terrestrial GW observatories currently operate. The jump to the kHz GW regime shows a much greater susceptibility to power loss by an exoplanet at the GW carrier frequency (due to the increase in the PMI by a factor of $2 \times 10^4$ caused by the change in the GW carrier frequency), suggesting that even potentially small exoplanets could produce notable effects.

Taking this one step further, we can use Equation (15) to compute the relative difference of the strain at the GW carrier frequency, with versus without the presence of an exoplanet:

$$\left.\frac{S^{(1)}_{H_a,\,\text{no exoplanet}} - S^{(1)}_{H_a,\,\text{exoplanet}}}{S^{(1)}_{H_a,\,\text{no exoplanet}}}\right|_{f = f_{\rm GW,0} + \frac{1}{2}\dot{f}_{\rm GW,0} T_{\rm obs}} \approx 1 - J_0^2(\epsilon). \quad (17)$$

Not only does this formula apply to the phase chirping signal, but this relationship also holds for pure monochromatic signals (i.e., in the limit $\dot{f}_{\rm GW,0} \to 0$).[25] Hence, Equation (17) gives us a way of predicting how different the peak output of power at the GW carrier frequency will be, with versus without the presence of an exoplanet.

Generally speaking, the presence of the exoplanet effectively behaves like a siphon. It draws power away from the peak of the GW carrier frequency and redistributes it into the sidebands at the integer multiples of the message frequency. The power loss at the GW carrier frequency is shown in Figure 3. The top panel of Figure 3 shows how the relative difference of the PSD at the carrier changes as a function of just the one-dimensional PMI parameter. We see that it is not until the PMI approaches $\epsilon \sim \mathcal{O}(1)$ before the loss of power at the GW carrier frequency becomes significant.

The PMI itself is dependent on GW carrier frequency, system mass, and the exoplanet orbital message frequency and mass (Equation (7)). Therefore, we can use Equation (17) as a proxy to help make predictions as to what part of the GW and the exoplanet parameter space this Doppler modulation observation/search might be the most sensitive to.

The bottom panels of Figure 3 show the exoplanet parameter space for two representative and fixed values of the GW frequency. We see that the predicted loss of power measured at the GW carrier due to an exoplanet in the LISA band will likely be very small, unless the exoplanet is a brown dwarf or even a low-mass star. This means that power in the exoplanet messages in the sidebands of the GW carrier will also be very small. Moreover, the LISA mission will have a nominal lifetime of 4 yr, with a planned extension to 10 yr, so even at longer exoplanet orbital periods (i.e., lower exoplanet orbital frequencies) the power drain may still only reach

[25] Indeed, one can even go through and rework all of the derivations carefully assuming no frequency evolution in the first place. In that case Equation (15) changes to $S^{(1)}_{H_a}(f = f_{\rm GW,0} + n f_p) \approx 2A^2 G_a J_n^2(\epsilon) T_{\rm obs}$. However, the resulting relative difference in the PSD is then still Equation (17).

$\mathcal{O}(<10\%)$. In the case of LISA and mHz GWs, we are likely in a small-$\epsilon$ regime.

However, we gain a remarkable insight from the dependence of the PMI on the carrier frequency. Since $\epsilon \propto f_{\mathrm{GW},0}$, going from the mHz to the kHz GW regime gives us a boost in the PMI by a factor $\mathcal{O}(10^4 - 10^6)$. We can understand this effect in connection to the physics behind the telecommunications analogy made in Section 1—the higher the carrier frequency, the more efficiently the message is transmitted. In this regime, we see that potentially even small, terrestrial-sized exoplanets could drain a significant amount of power from the GW carrier. In the case of the LVK and kHz GWs, we are likely in a large-$\epsilon$ regime.

The bottom two panels of Figure 3 also nicely demonstrate the GW RV method's sensitivity to high-mass, *long*-period exoplanets, as discussed in detail in Section 2.3. We also reemphasize the infinite exoplanet period caveat described at the end of Section 2.2. From these two plots it appears as if the power loss increases indefinitely as the orbital period goes to infinity. In reality, we need to take care in remembering that this proxy breaks down in certain limits. Namely, for real observations, the observation time of the experiment still will likely need to be several times longer than any underlying periodic signal in the data itself. Therefore, to detect the longer orbital period exoplanets, if they exist in these systems, an experiment will likely need to collect data for several times longer than $P$. This would, of course, place practical limitations on what can be measured.[26]

### 3.3. The Observer-frame Doppler Effect

The analysis so far has considered the observed spectrum of the GW system from a stationary observer frame. In reality, GW observatories are also in motion. For example, LISA will orbit the Sun with a 1 yr period (M. Colpi et al. 2024). This means that the LISA observatory itself may also move radially along the line of sight to any given GW source on the sky, over the course of a year. Therefore, LISA's own motion will induce a Doppler modulation on the observed signal, regardless of whether or not the GW source has an exoplanet.[27] In addition to its orbit, LISA will also "cartwheel" once per orbit, which adds further complications to modeling these relative Doppler effects.

All terrestrial GW observatories (such as the current LVK) experience similar effects. Since they are anchored to Earth, their motion also induces a 1 yr Doppler modulation on the signal due to Earth's orbital period. But in addition to that, Earth's own rotation adds a much shorter timescale Doppler modulation with its 1-day period. After an observation of only several hours, both of these Doppler modulations will become important when modeling long-lasting GW signals for these terrestrial GW observatories operating in the kHz GW spectrum (M. Maggiore 2007).

To build our understanding of how observer motion will complicate the overall observed GW spectrum, we focus here specifically on the LISA case. We anticipate that the results presented below will be similar for the LVK case. Interferometer GW observatories work on the premise that a GW will alter the light-travel time a photon takes to travel between two points, thanks to the strain induced (Equation (11)) on the surrounding spacetime. Therefore, in the presence of a GW, the net effect of shining a laser between two points is that the frequency of the laser light itself will fluctuate over time in an observable way.

LISA will function as a "synthetic" interferometer. Unlike the LVK observatory, which splits and recombines light in its interferometer, LISA will construct interferometric-like data through a post-processing procedure known as time-delay interferometry (TDI; M. Tinto et al. 2023). The TDI combinations can be made in many different ways—here we consider an example showing a "second-generation Michelson" combination (denoted below $X_{2.0}$). This $X_{2.0}$ observable itself physically represents a measurement of the laser fractional frequency fluctuation $\Delta f_{\mathrm{laser}}/f_{\mathrm{laser}}$ (hence, it is a unitless quantity).

For simplicity, we model the LISA constellation (i.e., the three spacecraft) using equal-arm-length orbits. The simulation of the signal in this work is achieved by combining the waveforms developed here (Equations (6), (11), and (12)) with some of the LISA Simulation Suite codes (see *Software* in the Acknowledgments). To compare the observed signal to the sensitivity of LISA under this assumption, we use the noise PSD model expressed in Equation (20) of S. Babak et al. (2021).[28]

Figure 4 demonstrates several examples of a LISA simulated observation of a binary system with an exoplanet. The left panel shows the full portion of the GW spectrum that LISA is sensitive to. These binary sources produce a sharp spectral line thanks to their monochromatic behavior. The top right panel zooms in on the signal in the left panel. Note that although the middle and bottom right panels correspond to two different sets of simulations, zooming out on them to the level of the full LISA spectrum would leave them qualitatively identical to the left panel as well.

We can gain insight into the relationship of the two Doppler effects (i.e., the one induced on the source by the exoplanet, and the other from the motion of the observer) by varying how long the exoplanet period $P$ is in relationship to the LISA period $T_{\mathrm{LISA}} = 1$ yr. For the first part of this discussion, let's just focus on the monochromatic signal in each of the panels.

The top right panel shows the case when $P \ll T_{\mathrm{LISA}}$, which means that $f_p \gg f_{\mathrm{LISA}}$. In this case, we still have the same general behavior described in Section 3.1, namely that the exoplanet message frequency appears attached to the GW carrier frequency at integer multiples in the sidebands.

However, the new addition we see here is the presence of a *second* message signal—the LISA orbital message. All of the closely spaced spikes in the top right panel correspond to $f_{\rm LISA} = 1/(1\ {\rm yr})$. This makes sense, because even without the exoplanet, the LISA orbit-induced Doppler FM should behave functionally in an identical way to what was discussed earlier for the exoplanet. Therefore, the LISA message appears in the sidebands around the GW carrier frequency at $n$ integer multiples: $f_{\rm GW,0} + nf_{\rm LISA}$. But it also appears in the sidebands around the exoplanet message frequencies. Now the exoplanet "messages" appear to act like frequency carriers to the LISA message: $f_p + nf_{\rm LISA}$. This behavior perhaps should not be surprising when again considering the telecommunications analogy: a message frequency can be attached to a carrier more efficiently when the carrier is much higher in frequency than the message. And that is exactly the scenario being shown in this panel.

[26] For example, if LISA observes for 4 (or 10) yr, then the practical limitation might be exoplanet periods of 2 (or 5) yr at most. For the LVK, that limitation may be similar or even longer, if the searches combine all observing run datasets that span its years/decade of observation. Future analyses will be needed to better understand these upper limits for exoplanet period detection.

[27] This is a demonstration of our classic understanding of Doppler physics—the effect is observed if the source is in motion (relative to the observer), if the observer is in motion (relative to the source), or both.

[28] This model also requires the noise PSD models of the optical metrology system ("OMS") and the spacecraft acceleration ("acc") that are given by Equations (10) and (13), respectively. In this work, we used slightly different values for the amplitudes of these PSDs, $A_{\mathrm{OMS}} = 12 \times 10^{-12}\ \mathrm{m}/\sqrt{\mathrm{Hz}}$ and $A_{\mathrm{acc}} = 2.4 \times 10^{-15}\ \mathrm{m/s^2}/\sqrt{\mathrm{Hz}}$.

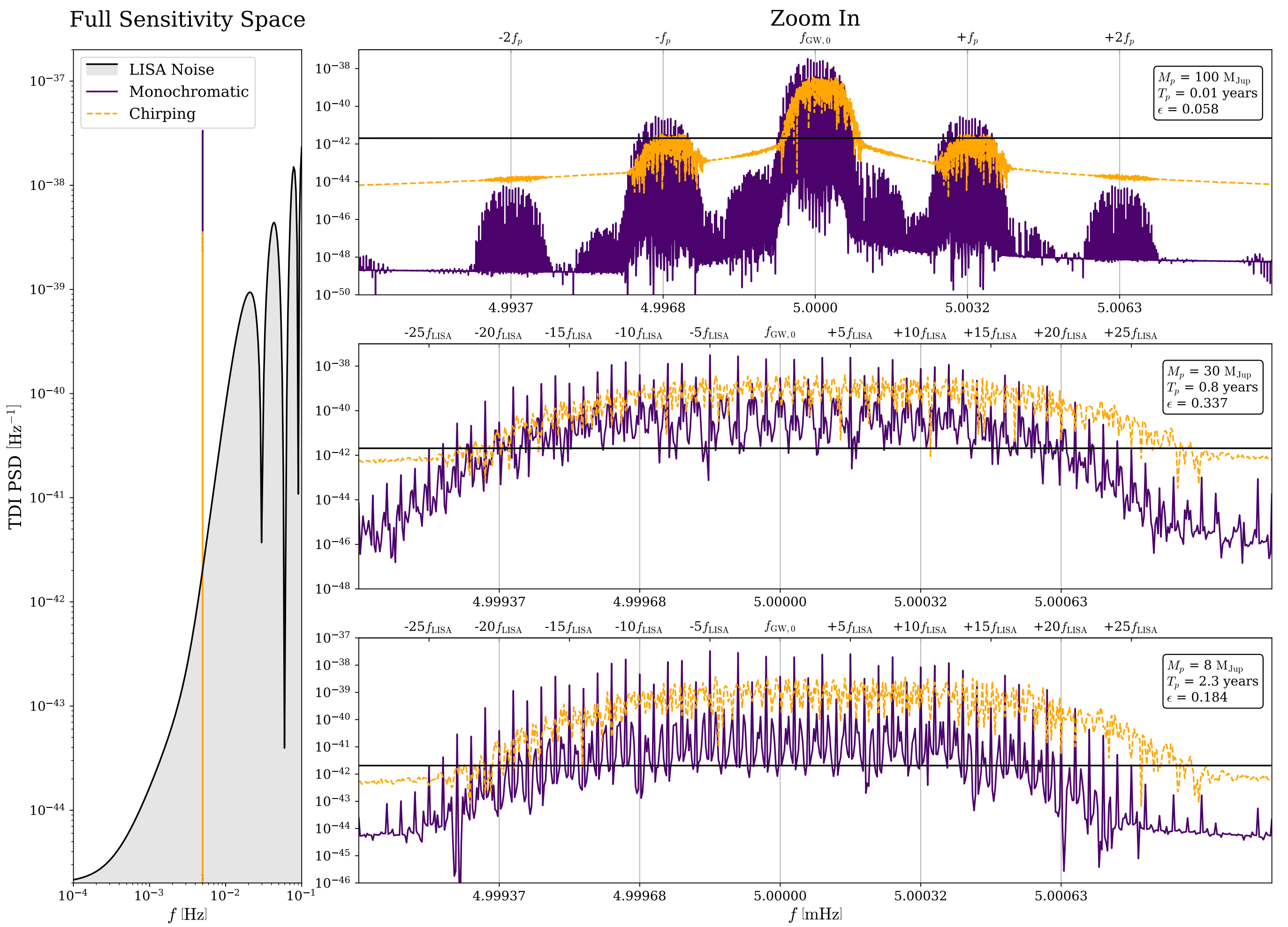

**Figure 4.** Observer and source Doppler effects for circumbinaries in LISA. Shown are side-by-side comparisons of the TDI PSD ($X_{2.0}$ Michelson combination) varying the length of the exoplanet period $P$ with relation to the duration of the LISA orbit $T_{\rm LISA}$. Here LISA is simulated with equal-arm-length orbits for a total observation $T_{\rm obs} = 10$ yr; therefore, the spectral frequency resolution is $\Delta f = 1/(10\ {\rm yr})$. For all simulations presented, $M_1 = M_2 = 0.6\,M_\odot$, $D = 2$ kpc, $i = \pi/2$ rad (maximal exoplanet Doppler effect), $\iota = \psi = \phi_0 = \phi_{0,p} = 0$ rad, the initial chirp rate is $\dot{f}_{\rm GW,0} = 22.62$ nHz yr$^{-1}$, and the total SNR is 567 (see Section 3.4). The source is at $\beta = \lambda = 0$ rad on the sky, which places it in the plane of the LISA orbit. This gives it the maximal observer Doppler effect. As the source changes its relative position on the sky to be perpendicular to the LISA orbit, the observer Doppler effect diminishes. Left: the part of the GW spectrum that LISA will be sensitive to, shown with an analytic model of the total noise sensitivity the mission is targeting. In all panels, the GW signal of the binary with the exoplanet is simulated at a carrier frequency $f_{\rm GW,0} = 5$ mHz. The monochromatic signals represent idealized signals (for comparison)—real GW signals are expected to chirp over time. The chirping simulated here is purely GW driven. Top right: $P \ll T_{\rm LISA}$. In this scenario, the GW frequency is a carrier to both the exoplanet and LISA messages, and the exoplanet message acts as an additional carrier to the LISA message. The ratio of the carrier frequency SNR and the $n = 1$ message frequency SNR is approximately 2.9%. Middle and bottom right: $P \sim T_{\rm LISA}$ and $P > T_{\rm LISA}$, respectively. In these cases, both LISA and exoplanet messages are carried by the GW carrier frequency and become highly mixed together. Using Equation (20), we analytically predict that the ratio of the carrier and $n = 1$ message SNRs is 17.1% and 9.2%, respectively. Notice that in these two plots we change the top $x$-axis to mark measures of $f_{\rm LISA}$, since the LISA messages are easier to distinguish by eye in the monochromatic signal.

The relationship between the LISA and exoplanet messages becomes trickier as the period of the exoplanet approaches the period of the LISA orbit, $P \sim T_{\rm LISA}$. An example of this case is shown in the middle right panel. Based on the understanding from the top right panel, we expect the two messages to be attached to the overall GW carrier frequency. The exoplanet message should appear at roughly the same spacing since its period is nearly the same, and indeed in the monochromatic case we see that there appear to be multiple sets of peaks interspersed in the sidebands, coming from the mixture of both LISA and exoplanet messages. Furthermore, we find from simulations that when the exoplanet period matches the LISA period, i.e., $P = T_{\rm LISA} = 1$ yr, we cannot distinguish the exoplanet's presence in the spectrum, because all of the messages align perfectly in the sidebands. Practically, it is unlikely that LISA will be able to find an exoplanet around a binary if the exoplanet's period is 1 yr.[29]

In the bottom right panel we look at the case when the exoplanet period is longer than the LISA period, $P > T_{\rm LISA}$. Qualitatively it does not appear much different from the middle right panel.[30] So in general, of these three cases, only if $P \ll T_{\rm LISA}$ can we expect a clear separation of the two sets of messages. Otherwise, both LISA and exoplanet messages will be present and mixed together in the sidebands around the GW carrier frequency.

Finally, in all cases we see that the effect of the frequency chirping on the signal is the same as described in Section 3.1, namely that it causes the spectrum of the signal to expand over time. Based on what we learned from the top panel of the binary column in Figure 2, we expect that each message is growing over time at the same rate as the carrier frequency. However, in Figure 4 this behavior is now difficult to distinguish, because now, with many more messages (both from the LISA orbit and the exoplanet) appearing in the sidebands, any growth over time $\Delta f_{\rm GW}$ quickly overlaps to create the intrasignal confusion in the observed spectrum described in Section 3.1 (just as we saw in the middle and bottom panels of the binary column).

### 3.4. Signal-to-noise Ratio Analysis

In any real experiment, the signal itself needs to be distinguishable from the underlying noise present in the GW instrument. The usual convention in GW astrophysics is to calculate the SNR of the observed signal to the expected noise floor, as given by the formula

$$\mathrm{SNR}^2 \equiv 4\ \mathrm{Re}\int_0^\infty \frac{|\tilde{M}(f)|^2}{S_N^{(1)}(f)}df, \tag{18}$$

where $S_N^{(1)}(f)$ is a model of the (one-sided) PSD of the instrument's noise and $\tilde{M}(f)$ is the signal model in the frequency domain. The full signal model is typically a more complicated linear combination of the two strain polarization states $M(t) \sim aH_+(t) + bH_\times(t)$ from Equation (11) that also encodes the specific geometry and properties of the GW instrument (C. J. Moore et al. 2015; S. Babak et al. 2021).

In order to make a simpler calculation just to build intuition and to remain agnostic to the detecting instrument (e.g., LISA vs. LVK, which have very different noise PSD models in different parts of the GW spectrum), we analytically solve for the SNR of $M = H_a$ just using the strain polarization function itself, Equation (11). Interestingly, just like with Equation (17), we find that the result is independent of whether or not the signal is purely monochromatic or chirping:[31]

$$\begin{aligned}\underbrace{\mathrm{SNR}^2_{H_a}}_{\text{total}} &\approx \sum_{n=-\infty}^{\infty}\underbrace{\mathrm{SNR}^2_{H_a,n}}_{n\text{th sideband}}\\ &\approx \sum_{n=-\infty}^{\infty}\frac{4A^2G_aJ_n^2(\epsilon)T_{\rm obs}}{S_N^{(1)}\left(f_{\rm GW,0}+nf_p+\frac{1}{2}\dot{f}_{\rm GW,0}T_{\rm obs}\right)},\\ &\approx \frac{4A^2G_aT_{\rm obs}}{S_N^{(1)}}.\end{aligned} \tag{19}$$

In general, for any of these sources (whether they are binaries or triaxial bodies), since they are so nearly monochromatic, the power spectrum of the instrument will be approximately constant over the range of frequencies that the sidebands cover. Therefore, we can approximate $S_N^{(1)}(f) \approx S_N^{(1)}$.

This result gives us several useful insights:

1. Thanks to the identity $\sum_{n=-\infty}^{\infty} J_n^2(x) \equiv 1$, this relationship tells us that the presence of an exoplanet does not change the *total* SNR of the observed system (compared to if it did not have an exoplanet). This further justifies the earlier statement made in Section 3.2 that the presence of the exoplanet redistributes power from the GW carrier frequency into the sidebands.[32] However, in practice, if only a subset of the total number of sidebands are correctly identified and included in this summation, then the total SNR may appear to decrease. This insight and its consequences are discussed further in Section 3.5.
2. Not only do we have a way of calculating the total SNR of the signal, but we can also calculate the individual SNR of just the GW carrier signal ($n = 0$) and each of the sidebands ($n \neq 0$). We even have a way of assessing how strong the message sidebands should be relative to the carrier:

$$\frac{\mathrm{SNR}(n^{\rm th}\ \text{sideband})}{\mathrm{SNR}(\text{GW carrier})} \approx \left|\frac{J_n(\epsilon)}{J_0(\epsilon)}\right|, \tag{20}$$

   which only depends on the PMI. What is also interesting about this formula is that because it is a *ratio* of the SNRs, it should still be relevant for the actual SNR of the observed signal when the previously mentioned geometry and properties of the GW instrument are correctly accounted for.
3. Equation (19) tells us that the SNR of *only* the GW carrier frequency ($n = 0$) itself drops by a factor of $|J_0(\epsilon)|$ compared to the same system that contains no exoplanet. A potential consequence of these insights is that if the exoplanet draws a significant amount of power away from the carrier but the sideband messages themselves are not observed or correctly modeled in the signal analysis (perhaps because the messages fall below the noise sensitivity curve of the instrument), then the signal may appear to have an incorrect SNR from

[29] Same for any terrestrial GW observatories, since they will have the same Doppler modulation.

[30] Practically, it does not really make sense to go to the extreme where $P \gg T_{\rm LISA}$, because past its nominal 4 yr mission LISA would be extended to 10 yr. And to observe a periodic signal, we still want the observation time to be at least a few times longer than the signal's period. But if we could increase the observation time to something much greater than 10 yr, then qualitatively we might expect the observed spectrum to look like the top right panel again, but with the exoplanet messages now tightly clustered around the GW carrier and the LISA message signals.

[31] We arrive at this conclusion by first discretizing Equation (18). The integral becomes a summation evaluated at each of the peaks in the PSD. In the monochromatic case, we let $df \to \Delta f = 1/T_{\rm obs}$, but in the chirping case, we let $df \to \Delta f_{\rm GW} = |\dot{f}_{\rm GW,0}|\ T_{\rm obs}$.

[32] We can think of this like a conservation law—the observed signal power is not removed, just redistributed.

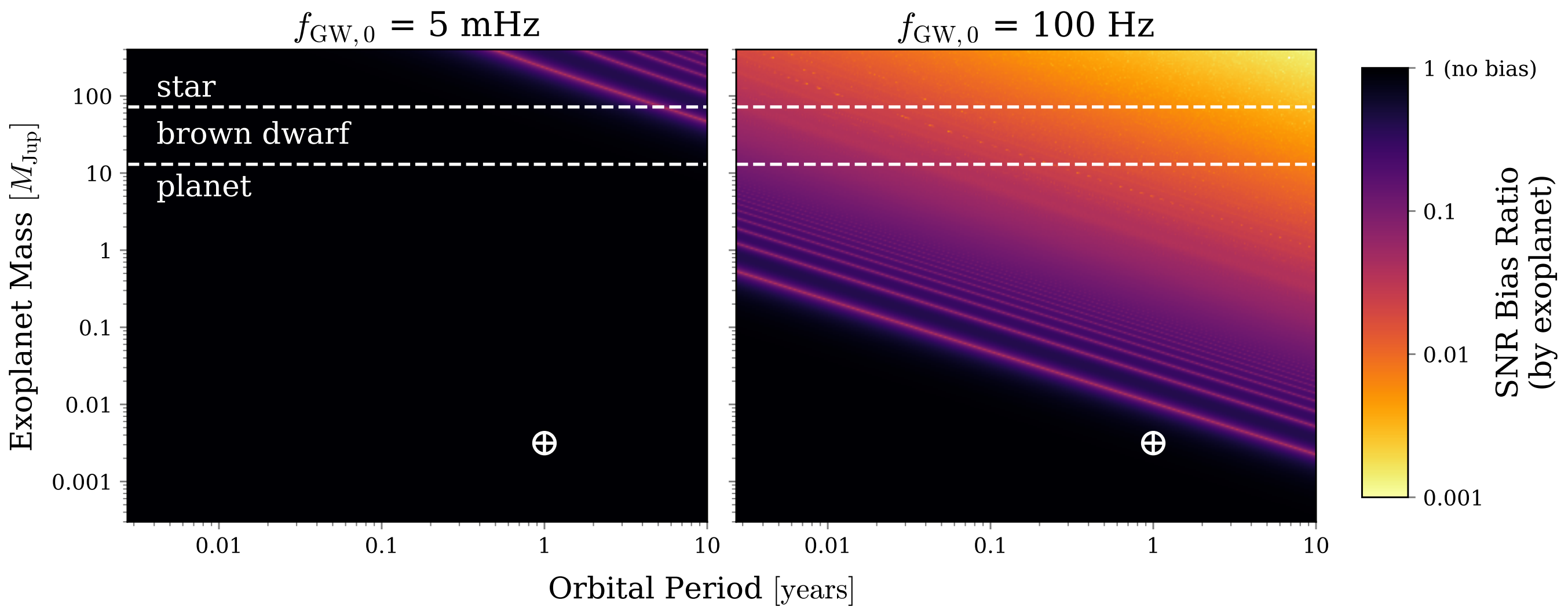

**Figure 5.** The exoplanet bias effect. Since the presence of an exoplanet removes power from the GW carrier frequency to supply the messages in the sidebands, the GW carrier itself will appear to have a lower SNR. If the signal is confused for being an isolated source with no exoplanet companion, then this may introduce a bias in other parameters of the GW waveform. Equation (21) gives us a way of quantifying this SNR bias ratio, $\mathcal{B}$. For example, if all parameters are held constant, then the bias could make it seem like the GW source is at a greater distance than it really is (for either the binary or triaxial body systems) by a factor of $D_{\rm biased} = \mathcal{B}^{-1}D$. Note that, just like in Figure 3, $M_{\rm sys} = 1.5\,M_\odot$, the exoplanet's inclination was set to $i = \pi/2$ (maximal Doppler effect), and the planetary/stellar mass boundaries are indicated. Left: a representation of the mHz GW regime, where the LISA mission will be most sensitive. Here we see that the effect of the source distance bias that an exoplanet may cause on the SNR of the measured signal may be mostly negligible across the discovery parameter space. Not until the mass of the "exoplanet" reaches the stellar mass range do we see bias factors of $\mathcal{B} \sim \mathcal{O}(0.1)$. Right: a representation of the kHz GW regime, where terrestrial GW observatories currently operate. Here we see a larger portion of the discovery parameter space where the bias could potentially cause the source to be confused for a distance of $>\mathcal{O}(10)$ farther away from Earth. Alternatively, the bias could make it seem like the neutron star has an even smaller ellipticity, $e_{\rm biased} = \mathcal{B}e$. Here we see the bias ratio $\mathcal{B} < \mathcal{O}(0.1)$ across much of the parameter space.

what it should. This is idea discussed in more detail in Section 3.5.

To see this in application for the LISA mission, we numerically calculate the SNRs of the signals shown in Figure 4. This numerical calculation therefore does include all of the proper modeling of the GW observatory geometry and motion and uses the appropriate PSD noise model. In all three examples, we calculate a total SNR of 567, regardless of whether the signal is monochromatic or chirping.[33] This reaffirms our analytic derivation of Equation (19). Next, we focus on the top right panel and begin with the SNR for the monochromatic signal. In this case, we calculate that the carrier SNR is 60.6 and the $n = 1$ message SNR is 1.7, meaning that the numerically calculated ratio of these is 2.8%. And indeed, we find that the analytic prediction of the ratio of these two SNRs calculated from Equation (20) is 2.9%. This analytical prediction is in agreement with the numerical result in spite of the fact that it was derived from Equation (19), which was agnostic to the GW observatory.

If we repeat these calculations for the top right panel but for the chirping signal, we find that the SNRs of just the carrier and just the $n = 1$ message numerically evaluate to values that strongly depend on the width of the frequency window around these frequencies for the integration. However, numerical *ratios* of these SNRs once again evaluate to approximately 2.9% and hence are still in agreement with the analytic prediction.

These comparisons between the numerical and analytic predictions work best in the case where $P \ll T_{\rm LISA}$. But in the bottom two panels, where the LISA orbital messages are mixed with the exoplanet messages and the intrasignal confusion is present, we find that we cannot numerically separate out the SNR of the exoplanet messages from the other effects as we did in the top right panel. In these cases, where $P \sim T_{\rm LISA}$ and $P > T_{\rm LISA}$, we rely on Equation (20) as a predictive diagnostic to understand how much of the signal's SNR is coming from the message versus the carrier, where the numerical calculation cannot give us insight. In these middle and bottom right panels, the analytically predicted ratios are 17.1% and 9.2%, respectively.

For a continuation of this study of the SNR with additional useful insights and application to the LISA mission, see C. McGrath et al. (2026).

### 3.5. Exoplanet Bias Effects

Since the presence of an exoplanet effectively siphons power to the sidebands of the GW signal, this may possibly introduce a form of bias into the estimation of some of the GW signal's parameters if the exoplanet is not correctly modeled and accounted for. We know from Section 3.4 that the *total* SNR of the observed signal does not change for a system with versus without an exoplanet. However, we also now understand that for systems that do contain an exoplanet, some of their signal's power gets removed from the GW carrier frequency and redistributed into the messages in the sidebands. Therefore, the peak of the PSD at the GW carrier frequency changes by a factor of $|J_0(\epsilon)|$ when it contains an exoplanet: $\mathrm{SNR}(\mathrm{f_{GW,0}}, \mathrm{exoplanet}) = |J_0(\epsilon)|\,\mathrm{SNR}(\mathrm{f_{GW,0}}, \mathrm{no\ exoplanet})$.

Mathematically, the SNR of a system without an exoplanet will change in a similar way by altering the signal's GW amplitude to a new amplitude: $\mathrm{SNR}(f_{\rm GW,0}, \text{no exoplanet, new}) = (A_{\rm new}/A)\ \mathrm{SNR}(f_{\rm GW,0}, \text{no exoplanet})$. So imagine a scenario where an observer detects a GW signal that contains an exoplanet and correctly identifies the peak at the carrier frequency but does not identify the sideband messages (perhaps, e.g., because these messages are buried below the noise in the data). Then, in this scenario, the observer may incorrectly interpret the GW system as having *less* power than it really does and believe that the system has no exoplanet.

[33] Although the monochromatic signal is louder at *specific* frequencies, the chirping signal is spread over a wider range of frequencies, so the total SNR integrated over the entire spectrum is the same.

This misinterpretation of setting the observed signal (with the exoplanet) to a new signal (without an exoplanet *and with a different amplitude*), that is, $\mathrm{SNR}(f_{\mathrm{GW},0}, \text{no exoplanet, new}) = \mathrm{SNR}(f_{\mathrm{GW},0}, \text{exoplanet})$, can be thought of as a bias ratio $\mathcal{B}$ on the GW amplitude and results in the following relationship:

$$\frac{A_{\mathrm{biased}}}{A} = |J_0(\epsilon)| \equiv \mathcal{B}. \tag{21}$$

Given the way it is defined here, $0 \leqslant \mathcal{B} \leqslant 1$, and a bias ratio $\mathcal{B} = 1$ indicates no bias.

Depending on the type of system (binary or triaxial body), the amplitude is given by either Equation (12) or Equation (14). From these two relationships, we see that the bias could manifest itself in different ways.[34] For example, holding all of the system parameters that appear in the PMI *constant* (Equation (7)), in both types of systems the bias could come from a bias in the source distance measurement: $D_{\mathrm{biased}} = \mathcal{B}^{-1}D$. For the triaxial body, the bias could also come from a bias in the ellipticity measurement: $e_{\mathrm{biased}} = \mathcal{B}e$. Here we do not focus on the potential biases in either the system chirp mass (Equation (13)) or the GW frequency parameters, because these parameters also appear in the PMI and hence their bias relationships are not as easy to define.

Figure 5 demonstrates this bias ratio. What is perhaps most interesting is that at the frequencies where the LISA mission will operate (left panel) the bias may be mostly negligible unless the companion reaches stellar masses (where we see bias factors $\mathcal{B} \sim \mathcal{O}(0.1)$). However, the implications for the terrestrial observatories might be significant. In addition to making the source potentially appear farther away, it could also confuse making the ellipticity of the neutron star appear more "spherical." Figure 5 suggests that across much of the shown parameter space the ellipticity could appear biased by factors $< \mathcal{O}(0.1 - 0.01)$.

## 4. Conclusion

*Caveats and discussion of triaxial body systems.* The primary goal of this work was to approach the study of the effect an exoplanet or companion might have on a GW system from a mostly agnostic point of view, by considering the two most likely GW source classes for this potential measurement possibility, but (with the exception of Section 3.3) without considering in detail the specific observing instrument. However, the sensitivity of the instrument itself and the strength of the GW signal ultimately will be critical for enabling such a detection and discovery.

From the analyses presented in Section 3, we see that the PMI itself, and consequently how the power is distributed in the observed spectrum, is strongly dependent on the GW carrier frequency. This, for example, gives Figure 3 a larger potential discovery space when looking at the relative difference in the strain of the GW carrier frequency (Equation (17)). However, with approximately a decade of searching, no detections of these triaxial systems have been made with the current generation of terrestrial observatories (see, e.g., R. Abbott et al. 2022; A. G. Abac et al. 2025a). The GW strain amplitude for triaxial bodies like neutron stars (Equation (14)) is generally expected to be much smaller than the strain amplitude for binary bodies like DWDs (Equation (12)). Therefore, the overall SNR of these triaxial systems is expected to be much smaller. This means that while Equation (17) is still useful for understanding how the signal power might be redistributed by an exoplanet, it by itself does not stand as a metric for "detectability"—that would require more analysis along the lines of what was presented in Section 3.4 that factors in observatory sensitivity.

Therefore, with the current lack of discovery of a triaxial body GW source, there is an intrinsic model degeneracy. It is difficult to determine whether the absence of a detectable signal is due to

1. a small neutron star ellipticity (since, by Equation (14), the GW strain is proportional to ellipticity $e$);
2. the presence of an unknown exoplanet (which, as shown in Section 3.5, could bias potential triaxial systems hosting exoplanets toward appearing even more spherical); or
3. other astrophysical processes that make detection more complicated.

Finally, while many thousands of pulsars have been discovered to date, pulsars discovered hosting exoplanets have been very rare, likely a consequence of the formation channels that could create such a system (R. G. Martin et al. 2016). However, a larger number of pulsars have been discovered hosting white dwarfs, main-sequence stars, and other objects (P. F. Wang et al. 2025). Therefore, while the prospect of discovering a planet around a GW system like this is exciting, the triaxial GW source system may not be the most likely candidate. If anything, a heavier object companion (like other white dwarfs or neutron stars) would produce a much greater PMI and therefore potential modulation effect on an observed GW signal.

*Summary.* The GW RV method is a relatively new idea for detecting exoplanets orbiting around GW sources. And while the method stems from the same approach as the traditional EM RV method, by inferring the presence of an exoplanet from Doppler modulations of the observed signal, the way the Doppler effect manifests itself in the GW signal is quite different—both conceptually and mathematically.

Conceptually, we provide an analogy for understanding GW RV from the physics behind telecommunications (Section 1). High, constant-frequency EM signals are excellent ways of sending low-frequency messages (such as audio that has been converted to EM) across great distances. For the GW signals in question, the sources are nearly monochromatic and much higher in frequency than the orbital frequency of the exoplanet. Therefore, in a very similar way to the carrier and message frequencies of telecommunications, the GW source behaves like a carrier signal to the exoplanet's orbital message. The resulting observed spectrum in GW RV therefore appears like the observed spectrum of a telecommunications signal.

Mathematically, in this work we provide a more detailed derivation of the Doppler effect on the observed GW frequency and phase for systems in circular orbits than shown in previous work (Section 2). Crucially, we derive the analytic formula for the observed GW phase and point out the significance of the PMI $\epsilon$. This quantity (not discussed in previous GW RV studies), while similar to the semiamplitude parameter for traditional EM RV, is the relevant parameter that controls the strength of the GW RV exoplanet modulation. And we find that this parameter is more sensitive to high-mass, *long-period* exoplanets for GW RV. This represents a significant departure from our traditional thinking of EM RV, which is more sensitive to low-period exoplanets (this comparison between EM and RV methods is made in Section 2.3).

[34] In the limit that the exoplanet vanishes, i.e., $\epsilon \to 0$, the bias $\mathcal{B} \to 1$ and disappears as we would expect.

We discuss the two primary sources for consideration of GW RV: binary systems and triaxial bodies. Binary GW systems would likely consist of ultracompact stars in combinations of white dwarfs, neutron stars, and black holes falling in the mid-frequency GW spectrum, where future missions like LISA will operate. Most attention to date on the GW RV method has been focused on these systems, and in particular DWDs, since they will be the single most common type of signal for LISA. However, we add to the discussion triaxial bodies, with the most likely system being isolated asymmetric neutron stars falling in the high-frequency GW spectrum, where the current terrestrial LVK GW observatories operate.

With the signal model and these types of systems in mind, we perform several frequency domain analyses to better understand the fundamental physics behind the observed GW spectrum for a source hosting an exoplanet (Section 3). For simplicity, we begin by performing analyses in just the GW system reference frame. If the system is truly monochromatic, we see the analogy to a telecommunications spectrum in action with the GW carrier frequency surrounded by exoplanet message frequencies attached to it in the sidebands. For signals with frequency chirping, this evolution over the observation window can cause intrasignal confusion that can make the sideband message difficult to distinguish from the carrier.

But in spite of this, we are able to analytically derive a simple formula for the observed PSD of the GW strain. Going one step further, we find that the relative difference of the GW strain at the GW carrier frequency is *independent* of whether or not the underlying system is truly monochromatic or has some frequency chirp. We present this key finding, a result that is only a function of the PMI, as one way to map out the potential exoplanet discovery space. And we find that in the mid-frequency GW spectrum the "exoplanet" systems may have to be significantly massive (such as brown dwarfs or even stellar companions), in order to siphon a significant amount of power from the GW carrier frequency to supply the exoplanet messages in the sidebands. But just like with telecommunications—the higher the carrier frequency, the easier it is to send the message—we find that in the high-frequency GW spectrum even relatively small exoplanets (terrestrial sized) may result in a significant power siphon to the GW signal. We show that, thanks to the behavior of the PMI, going from the mHz GW spectrum to the kHz GW spectrum increases the PMI by $\sim\mathcal{O}(10^4 - 10^6)$.

We also present a more focused analysis of the additional Doppler effect that is introduced when considering the observer reference frame, specifically modeling the upcoming LISA observatory. Since LISA will orbit the Sun with a period of 1 yr, the observatory's motion introduces a Doppler effect as well, which attaches a *second* message frequency to the carrier—the LISA orbital message. We find that when an exoplanet's period is much less than the LISA orbital period, we have a clear separation of effects: the exoplanet message frequency attaches to the GW carrier frequency as we found before, and additionally the LISA orbital frequency attaches itself to *both* the GW carrier and each exoplanet message. As the exoplanet and LISA periods become more similar in duration, these two messages become highly mixed with each other.

Then, we present an analysis of how the SNR changes with the presence of an exoplanet. We do acknowledge that the calculation is somewhat simplified. Rather than analytically solve the SNR using a complete signal model for a specific GW observatory (which would require the geometry and response of the specific GW detector of interest), we instead remain mostly agnostic of instrument and use as a proxy only the GW strain model. Nevertheless, this gives us several key findings. Perhaps most usefully, we derive an analytic relationship that lets us calculate the ratio of the SNR of the GW carrier to the SNRs of the exoplanet messages in the sideband, and it again only depends on the PMI. This gives us another useful predictive tool to quickly calculate the expected strength of an exoplanet on the overall observed signal, without the need for complex modeling. This study is continued in C. McGrath et al. (2026).

We also investigate how the presence of an exoplanet, if undetected, might lead to a bias effect in the estimation of certain GW system parameters. Specifically, we find that in the mid-frequency GW spectrum the bias in GW source distance parameter may be mostly negligible across the exoplanet discovery space unless the companion reaches stellar masses. It is worth pointing out, however, that this was a relatively simple analysis. M. L. Katz et al. (2022) also discuss the potential for bias within the context of undetected companions orbiting DWD systems observed by LISA, which they found to be more significant in their more in-depth Bayesian analysis. In the high-frequency GW spectrum, we find that the potential for bias could be more significant. An unmodeled exoplanet might make the source appear farther away than it really is. It could also make the potential neutron star in question appear more spherical than it really is, reducing its ellipticity by factors $<\mathcal{O}(0.1 - 0.01)$ or more. As the LVK searches for GW signals from unknown neutron stars, careful modeling of the source signal and anything that might be modulating it becomes especially important since they lack complimentary EM observations (R. Abbott et al. 2022). Therefore, considering the possibility that undiscovered exoplanets could bias the ellipticity parameters to even lower values than expected in these types of systems could have serious consequences.

Lastly, in the appendices we provide analytic formulae for the frequency domain solutions of both purely monochromatic and phase chirping, circular orbit, GW systems with exoplanets. These may be useful tools in the development of future GW models and search pipelines.

*Final discussion.* In this work we considered searching for exoplanets that come from systems that generate GWs in the mid-frequency (mHz) and high-frequency (kHz) parts of the GW spectrum. This was primarily because there are both known candidates in these regimes and currently operating or future planned GW observatories sensitive to these regimes. But there are also the low-frequency (nHz), the $\mu$Hz, and the dHz parts of the GW spectrum. One of the key assumptions we made was that the GW carrier frequency was greater than the exoplanet message frequency (recall, too, the telecommunications analogy). Therefore, in the lower-frequency regimes (nHz and $\mu$Hz) the exoplanet periods would have to be very long, which observationally may make the detection difficult.[35] And the main candidates of GW sources in these regimes are supermassive black hole binaries, so the "exoplanets" may have to be third-companion supermassive black holes in a circumbinary with the GW source to produce a signal like this.

Perhaps more favorable would be the dHz regime, which would still contain many of the same types of the binary systems as found in the mHz GW regime, but whose frequencies are higher (M. Arca Sedda et al. 2020). This possibility has been investigated specifically in the context of the DECIGO mission concept (W.-L. Guo et al. 2025). Additionally, as we understand now from the PMI's properties, this regime would potentially benefit in a $\sim\mathcal{O}(10^2)$ boost in the PMI as compared to the mHz regime, making this an intriguing consideration for future study.

Future, more complex and robust analyses will be needed to better understand the ability to detect exoplanets in GW signals. But hopefully this work can better shed light on the fundamental physics that can inform those studies. There are numerous reasons why a GW RV method could be a great new tool for exoplanetary discovery. For one, GW observatories have potentially much farther reach into the Milky Way and beyond, since the GW amplitude is inversely proportional to the source distance. They will also inherently be probing post-main-sequence stellar systems, where EM observatories simply have not found as many exoplanets.

And if such systems were observed in both EM and GW, then they could become the first discovered multimessenger exoplanets (see, e.g., C. McGrath et al. 2026). However, such a discovery would not be without its challenges. GW observations typically have poor sky localization compared to EM observations. Additionally, objects such as white dwarfs and neutron stars are typically much fainter electromagnetically, due to their small size. Nevertheless, with the continued growth and expansion of GW astrophysics, opportunities and possibilities such as this should continue to be explored.

## Acknowledgments

The work presented here took a couple of years to develop, and over that time I benefited from many very helpful conversations with colleagues. Some introduced me to new topics, some suggested new ideas, some let me talk through my work in progress while giving feedback, and some offered helpful guiding questions, as well as answers to my own questions. I would like to thank the following individuals for the help and support that they provided me: James Ira Thorpe, Scott Noble, Jeremy Schnittman, Richard Barry, John Baker, Nicola Tamanini, Camilla Danielski, Michael Katz, Jean-Baptiste Bayle, Eleonora Castelli, Martin Staab, Olaf Hartwig, Cecilia Chirenti, Zachary Metzler, Ravi Kopparapu, Mary Anne Limbach, Vincent Kofman, Joe Renaud, and Emily Gilbert. I would also like to thank in general the LISA Simulation Working Group and the LISA Simulation Expert Group, for their discussions on all simulation-related activities, and the anonymous reviewer who provided really excellent insights, feedback, and suggestions when editing this paper.

I would especially like to credit Cecilia Chirenti for the idea to extend this work to the possibility of detecting exoplanets around neutron stars (triaxial bodies). I had initially focused only on the LISA and binary systems case, but this suggestion greatly expanded the scope and significance of the work.

The material presented here is based on work supported by NASA under award No. 80GSFC24M0006. I also acknowledge support from the NASA Goddard Space Flight Center "Sellers Exoplanet Environments Collaboration" (SEEC), which is funded by the NASA Planetary Science Divisions Internal Scientist Funding Model, and from a Goddard "Science Task Group" award.

*Materials Availability:* All code and materials to reproduce the calculations and the plots in this paper are publicly available under an open-source Apache-2.0 License: doi: 10.5281/zenodo.22129575.

*Software:* NumPy (C. R. Harris et al. 2020), SciPy (P. Virtanen et al. 2020), Matplotlib (The Matplotlib Development Team 2025), LISAOrbits (J.-B. Bayle et al. 2024b), LISAGWResponse (J.-B. Bayle et al. 2024a), PyTDI (M. Staab et al. 2023).

## Appendix A
## GW Parameter Degeneracies and Constraints

For a binary system, the source's distance and chirp mass are degenerate with each other in the combination of $\mathcal{M}^{5/3}/D$ in the GW amplitude. And in the triaxial body system, the source's distance, ellipticity, and mass quadrupole are all degenerate with each other in the combination $eI_3/D$. However, if we can measure the GW amplitude, frequency $f_{\rm GW,0}$, and chirp $\dot{f}_{\rm GW,0}$, then in theory some of these degeneracies can be broken, and the additional source parameters may be recovered or constrained.

If the system dynamics of either the binary or the triaxial body are governed *purely* by GWs, then their frequency evolutions have known analytical expressions:

$$\text{Binary System:} \quad \dot{\omega}(t) = \frac{3}{8}\frac{\omega_0}{\Delta\tau_c}\left[1 - \frac{(t-t_0)}{\Delta\tau_c}\right]^{-11/8}, \tag{A1}$$

$$\text{where } \Delta\tau_c \equiv \frac{5}{256}\left(\frac{c^3}{G\mathcal{M}}\right)^{5/3}\frac{1}{\omega_0^{8/3}},$$

$$\text{Triaxial Body:} \quad \dot{\omega}(t) = -\frac{1}{4}\frac{\omega_0}{\Delta\tau}\left[1 + \frac{(t-t_0)}{\Delta\tau}\right]^{-1/4}, \tag{A2}$$

$$\text{where } \Delta\tau \equiv \frac{5}{128}\left(\frac{c^5}{Ge^2I_3}\right)\frac{1}{\omega_0^4}.$$

Binaries have a spin-up chirp, while triaxial bodies have a spin-down chirp. The quantity $\Delta\tau_c$ is the time of coalescence (measured from the initial time $t_0$) of the binary. For the assumptions of this work, the coalescence time of the binary is much greater than the observation time of the experiment, $\Delta\tau_c \gg T_{\rm obs}$. The dynamical time parameter $\Delta\tau$ for the triaxial body is analogous to an "inverse" coalescence time (running a clock backward would evolve the system to something that mimics a "coalescence," where the body has infinite angular frequency).

Solving Equation (A1) at time $t = t_0$ for the binary lets us fully break the distance and chirp mass degeneracy. Unfortunately, for the triaxial body, solving Equation (A2) at time $t = t_0$ does not fully break the distance, ellipticity, and mass

[35] For example, pulsar timing techniques (that search the nHz GW spectrum) have collected over a decade of data and have not yet found a continuous GW source. Recall also our earlier discussion about the importance of observing long enough to catch multiple periods of the potential exoplanet.

quadrupole degeneracy, but it does allow us to place further constraints on these parameters:

$$\text{Binary System}\begin{cases} \mathcal{M} = \frac{c^3}{G}\left[\frac{5}{96\pi^{8/3}}\frac{\dot{f}_{\rm GW,0}}{f_{\rm GW,0}^{11/3}}\right]^{3/5}, \\ D = \frac{5c}{48\pi^2}\frac{\dot{f}_{\rm GW,0}}{A_{\rm binary}f_{\rm GW,0}^3}, \end{cases} \tag{A3}$$

$$\text{Triaxial Body}\begin{cases} e^2 I_3 = -\frac{5c^5}{32\pi^4 G}\frac{\dot{f}_{\rm GW,0}}{f_{\rm GW,0}^5}, \\ eD = -\frac{5c}{16\pi^2}\frac{\dot{f}_{\rm GW,0}}{A_{\rm triax}f_{\rm GW,0}^3}. \end{cases} \tag{A4}$$

However, if these systems contain other driving orbital dynamics in addition to GW radiation, then the spin-up/spin-down rates may differ from what is expected from gravitational radiation alone. For example, accreting environments around binaries could cause them to *spin down* (K. Kremer et al. 2017), and pulsars are known to undergo rapid sudden *spin-up* "glitch" evolutions (D. R. Lorimer & M. Kramer 2012; D. Antonopoulou et al. 2022). If dynamical evolutions such as these also exist in the system, then Equations (A3) and (A4) will no longer hold. But the relationships may still be useful for setting limiting constraints on parameters.

## Appendix B
## Explanation of Derivations

*Conventions.* To perform the frequency domain analysis, we need to take a Fourier transform of our time-domain signal. Instead of making the idealized assumption that the signal is periodic on an infinite time domain, we make the calculation assuming a finite observation window $T$. This enables us to mathematically derive how the signal changes and grows with time.

For a signal $M(t)$ observed beginning at time $t_0$, we can calculate the signal's Fourier transform and its one-sided PSD using the following relationships:

$$\mathcal{F}[M(t)] \equiv \tilde{M}(f) \equiv \int_{t_0}^{t_0+T} M(t)e^{-i2\pi ft}dt \quad \text{on the interval: } t_0 \leqslant t \leqslant t_0 + T, \tag{B1}$$

$$= e^{-i2\pi ft_0}\int_0^T M(t+t_0)e^{-i2\pi ft}dt \quad \text{on the interval: } 0 \leqslant t \leqslant T, \tag{B2}$$

$$= e^{-i2\pi f\left(t_0+\frac{T}{2}\right)}\int_{-\frac{T}{2}}^{\frac{T}{2}} M\left(t+t_0+\frac{T}{2}\right)e^{-i2\pi ft}dt \quad \text{on the interval: } -\frac{T}{2} \leqslant t \leqslant \frac{T}{2}, \tag{B3}$$

$$S_M^{(1)}(f) \equiv \lim_{T\to\infty}\frac{2}{T}|\tilde{M}(f)|^2. \tag{B4}$$

While the Fourier transform formulae are each equivalent, sometimes choosing one over another can aid in the derivation. As an example, we chose to derive the monochromatic formula (Equation (C3)) using Equation (B3) and the chirping formulae (Equations (C6) and (C9)) using Equation (B2).

Another benefit of writing these formulae this way is that discretizing them gives us the appropriate conversion between the Fourier transform and the discrete Fourier transform (DFT). Namely, for a finite array of $k$ frequencies $f_k \equiv k\,\Delta f$ (where $\Delta f = 1/T$ is the frequency resolution), the relationship between the Fourier transform $\tilde{M}(f)$ and the DFT $\tilde{M}_{\rm DFT}(f_k)$ is

$$\tilde{M}(f_k) = e^{-i2\pi f_k t_0}\frac{\Delta t}{G}\tilde{M}_{\rm DFT}(f_k), \tag{B5}$$

where $\Delta t$ is the spacing between the finite time samples $t_n = t_0 + n\Delta t$. Here $G$ is the normalization convention that is used in the DFT.[33] Equation (B5) gives us a convenient way of checking the validity of the analytically derived formulae with numerical tests. We can calculate the DFT from the time-domain version of our signal and then compare that to the output of the signal's analytic Fourier transform at those same frequencies.

*Helpful relationships.* The following relationships are useful for deriving the Fourier transforms and PSDs given in Appendix C:

$$e^{\pm i\epsilon\sin(\phi_p+\omega_m t)} = \sum_{n=-\infty}^{\infty} e^{in\phi_p}(\pm 1)^n J_n(\epsilon)e^{in\omega_m t},$$

$$C(-x) = -C(x) \quad \text{and} \quad S(-x) = -S(x),$$

$$\int_{-\frac{T}{2}}^{\frac{T}{2}} e^{iat}dt = \frac{2\sin\left(a\frac{T}{2}\right)}{a},$$

$$\lim_{a\to 0}\frac{\sin\left(a\frac{T}{2}\right)}{a} = \frac{T}{2},$$

$$\int_a^b e^{\pm i\frac{\pi}{2}x^2}dx = [C(b) - C(a)] \pm i[S(b) - S(a)],$$

$$\lim_{x\to\infty} C(x) = \lim_{x\to\infty} S(x) = \frac{1}{2},$$

where $J_n(x)$ is the $n$th Bessel function of the first kind and $S(x)$ and $C(x)$ are the Fresnel integrals.

## Appendix C
## Frequency Analysis

The signals $M(t)$ are defined in the time domain such that

$$M(t) = A\cos(\Phi(t)) \text{ on the interval: } t_0 \leqslant t \leqslant t_0 + T, \tag{C1}$$

where $\Phi(t)$ defines the phase of the signal. Figures C1 and C2 provide helpful visual references and breakdowns of many of the spectral features and terminologies of the signals as they are mathematically derived in the following sections.

### *C.1. Monochromatic Signal + Frequency Modulation*

A monochromatic signal oscillates at a constant carrier frequency $\omega_c$. For a (constant-amplitude) signal whose frequency is modulated at the message frequency $\omega_m$, such that $\omega_c > \omega_m$, the phase of the signal can be written as

$$\Phi(t) = \phi_0 + \omega_c(t - t_0) + \epsilon\sin(\phi_p + \omega_m(t - t_0)), \tag{C2}$$

where $\phi_0$ and $\phi_p$ are additional phase factors and $\epsilon$ controls the strength of the modulation, the PMI.

The Fourier transform of this signal is

[33] For example, the Python NumPy FFT package currently uses $G = \{1,\ 1/N,\ 1/\sqrt{N}\}$ for the "backward," "forward," and "ortho" normalization conventions, respectively.

$$\tilde{M}(f) = A \sum_{n=-\infty}^{\infty} J_n(\epsilon) \Bigg[ \overbrace{e^{i\left(\phi_0 + n\phi_p - \omega t_0 + (\omega_c + n\omega_m - \omega)\frac{T}{2}\right)} \frac{\sin\left((\omega_c + n\omega_m - \omega)\frac{T}{2}\right)}{(\omega_c + n\omega_m - \omega)}}^{\text{positive frequencies}} + \underbrace{(-1)^n e^{i\left(-\phi_0 + n\phi_p - \omega t_0 + (-\omega_c + n\omega_m - \omega)\frac{T}{2}\right)} \frac{\sin\left((-\omega_c + n\omega_m - \omega)\frac{T}{2}\right)}{(-\omega_c + n\omega_m - \omega)}}_{\text{negative frequencies}} \Bigg]. \tag{C3}$$

Recall that $\omega \equiv 2\pi f$ (so we can go back and forth between $\omega$ and $f$ when it is convenient). Notice the behavior of this function. We see that there are two primary sets of "delta-like" functions, with peaks at $\omega = \pm\,\omega_c + n\omega_m$ (i.e., $f = \pm f_c + nf_m$). For our problem of interest, the negative frequencies have no physical significance. And in many cases these terms can be omitted without losing accuracy.

We can calculate the PSD at each of these peaks to find the peak power at these special frequencies. To do so, we first carefully evaluate the $\tilde{M}(f)$ at the peaks and then use Equation (B4) to find

$$S_M^{(1)}(f = f_c + nf_m) \approx \frac{A^2 J_n^2(\epsilon)}{2} T. \tag{C4}$$

Therefore, the power spectrum behaves like a delta function with peaks at the carrier frequency and in the sidebands around the carrier frequency at integer values of the message frequency. And due to the limit definition of Equation (B4), these formulae for the peak values (Equations (C4), (C7), and (C10)) improve in accuracy as the observation window grows larger. Examples of monochromatic, FM signals are shown in panels (a), (b), and (c) of Figure C1.

### C.2. Chirping Signal (Spin-up) + Frequency Modulation

A chirping signal evolves away from the initial carrier frequency. If the evolution tends toward higher frequencies, $\dot{\omega}_0' > 0$, the signal has a spin-up chirp:

$$\Phi(t) = \phi_0 + \omega_c(t - t_0) + \epsilon \sin(\phi_p + \omega_m(t - t_0)) + \frac{1}{2}\dot{\omega}_0'(t - t_0)^2. \tag{C5}$$

The Fourier transform of this (constant-amplitude) signal is

$$\begin{aligned} \tilde{M}(f) = e^{-i\omega t_0}\frac{A}{2}\sqrt{\frac{\pi}{\dot{\omega}_0'}} \sum_{n=-\infty}^{\infty} e^{in\phi_p} J_n(\epsilon)[e^{i(\phi_0 - \alpha_-^2)} \\ \times ([C(X_u^-) - C(X_l^-)] + i[S(X_u^-) - S(X_l^-)]) \\ + (-1)^n e^{-i(\phi_0 - \alpha_+^2)}([C(X_u^+) - C(X_l^+)] \\ - i[S(X_u^+) - S(X_l^+)])], \end{aligned}$$

$$\text{where} \quad \alpha_\mp \equiv \frac{(\omega_c \pm n\omega_m \mp \omega)}{\sqrt{2\dot{\omega}_0'}}, \quad X_l^\mp \equiv \frac{(\omega_c \pm n\omega_m \mp \omega)}{\sqrt{\pi\dot{\omega}_0'}}, \quad X_u^\mp \equiv \sqrt{\frac{\dot{\omega}_0'}{\pi}}T + X_l^\mp. \tag{C6}$$

We see another separation of the positive and negative frequency terms as we did in Equation (C3). However, now with frequency chirping, the "peaks" in the PSD that we saw in the monochromatic case will spread out with observation time, between $\omega = \omega_c + n\omega_m \to \omega_c + n\omega_m + \dot{\omega}_0' T$. Therefore, let's treat $\omega = \omega_c + n\omega_m + \frac{1}{2}\dot{\omega}_0' T$ (i.e., $f = f_c + nf_m + \frac{1}{2}\dot{f}_0' T$) as the representative "center" of every peak in this new power spectrum as it spreads with observation time (see, e.g., the upper right corner of panel (c) in Figure C1).

Once again, we can calculate the PSD at each of these representative peaks by carefully evaluating Equation (C6) in the limit that $\omega \to \omega_c + n\omega_m + \frac{1}{2}\dot{\omega}_0' T$ and then using Equation (B4) to find

$$S_M^{(1)}\left(f = f_c + nf_m + \frac{1}{2}\dot{f}_0' T\right) \approx \frac{A^2 J_n^2(\epsilon)}{2\dot{f}_0' T}. \tag{C7}$$

Examples of chirping, FM signals are shown in panel (c) (bottom right) of Figure C1 and Figure C2.

### C.3. Chirping Signal (Spin-down) + Frequency Modulation

If, instead, the frequency evolution of the signal tends toward lower frequencies, $\dot{\omega}_0' < 0$, then the signal has a spin-down chirp:

$$\Phi(t) = \phi_0 + \omega_c(t - t_0) + \epsilon \sin(\phi_p + \omega_m(t - t_0)) - \frac{1}{2}|\dot{\omega}_0'|\,(t - t_0)^2. \tag{C8}$$

The Fourier transform of this (constant-amplitude) signal is nearly identical to the spin-up case, but with some key sign changes:

$$\begin{aligned} \tilde{M}(f) = e^{-i\omega t_0}\frac{A}{2}\sqrt{\frac{\pi}{2|\dot{\omega}_0'|}} \sum_{n=-\infty}^{\infty} e^{in\phi_p} J_n(\epsilon)[e^{i(\phi_0 + \alpha_-^2)} \\ \times ([C(X_u^-) - C(X_l^-)] - i[S(X_u^-) - S(X_l^-)]) \\ + (-1)^n e^{-i(\phi_0 + \alpha_+^2)}([C(X_u^+) - C(X_l^+)] \\ + i[S(X_u^+) - S(X_l^+)])], \end{aligned}$$

$$\text{where} \quad \alpha_\mp \equiv \frac{(\omega_c \pm n\omega_m \mp \omega)}{\sqrt{2|\dot{\omega}_0'|}}, \quad X_l^\mp \equiv -\frac{(\omega_c \pm n\omega_m \mp \omega)}{\sqrt{\pi|\dot{\omega}_0'|}}, \quad X_u^\mp \equiv \sqrt{\frac{|\dot{\omega}_0'|}{\pi}}T + X_l^\mp. \tag{C9}$$

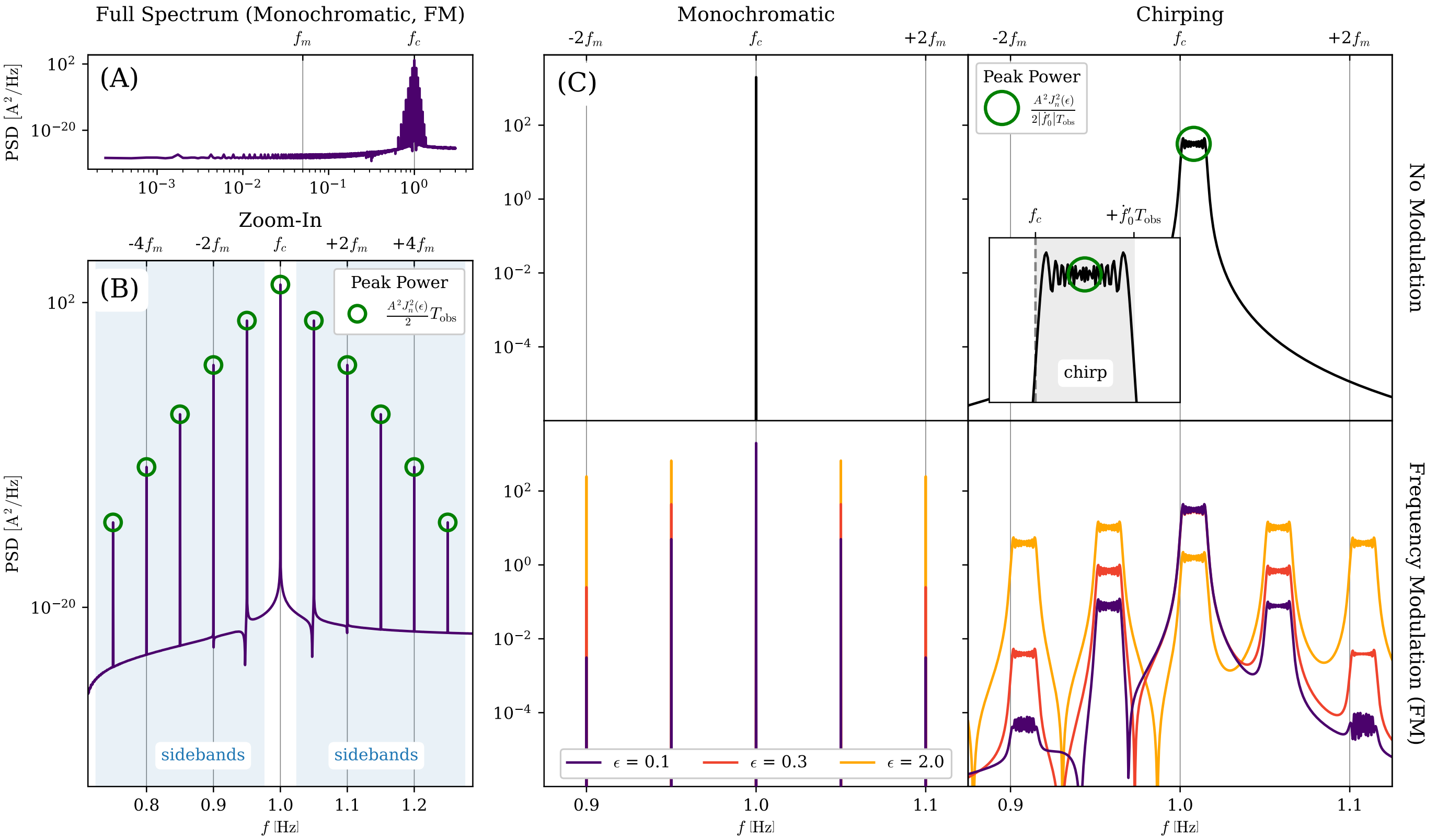

**Figure C1.** Identifying features in an FM signal's spectrum. Here we highlight key features of the signal spectrum of an FM signal, including when it is a purely monochromatic signal and when there is some frequency chirping present. Signals of the same color but used in different panels of this mosaic plot are the same signal (just shown with different axis scales). For every plot in this mosaic, the axes are frequency (bottom axis, in hertz) versus PSD (side axis, in squared amplitude units per hertz). (a) The full spectrum of a monochromatic FM signal, with carrier frequency $f_c$ and message frequency $f_m$ indicated along the top axis. Recall that for FM signals $f_m < f_c$. All spectral information in this signal is centered about the carrier frequency. (b) The same signal as in panel (a), zoomed in to the frequencies with spectral information. The sidebands (the regions of frequency space immediately surrounding the carrier frequency) for an FM signal contain the message frequency at $n$ integer multiples (indicated along the top axis) and are highlighted here. Additionally, the peak power as calculated in Equation (C4) is indicated. (c) A side-by-side comparison showing how the signal changes between monochromatic and chirping signals, and signals with FM vs. without any modulation. Top left: the simplest case—a monochromatic signal without any modulation just looks like a delta function. Top right: adding frequency evolution to the signal $\dot{f}_0'$ causes that signal to chirp over the observation time $T_{\rm obs}$. This changes where we choose to calculate the representative "peak" in the power spectrum—i.e., at the midpoint $\frac{1}{2}\dot{f}_0' T_{\rm obs}$—as given by Equation (C10). This example shows a positive chirp (or "spin-up"), starting at the carrier and evolving toward higher frequencies. Bottom left: several monochromatic FM signals are plotted with increasing values of the PMI $\epsilon$. As $\epsilon$ grows, so does the amount of power contained in each sideband message. Bottom right: the same three signals, but now with chirping. Visible here (and present but not visible in the bottom left panel), as the power grows in the sidebands, the power in the carrier frequency decreases. In more extreme cases, as $\epsilon$ grows to values >1, the peak in power at the carrier frequency itself can actually be *lower* than the power in the sideband messages (seen here in the $\epsilon = 2$ case).

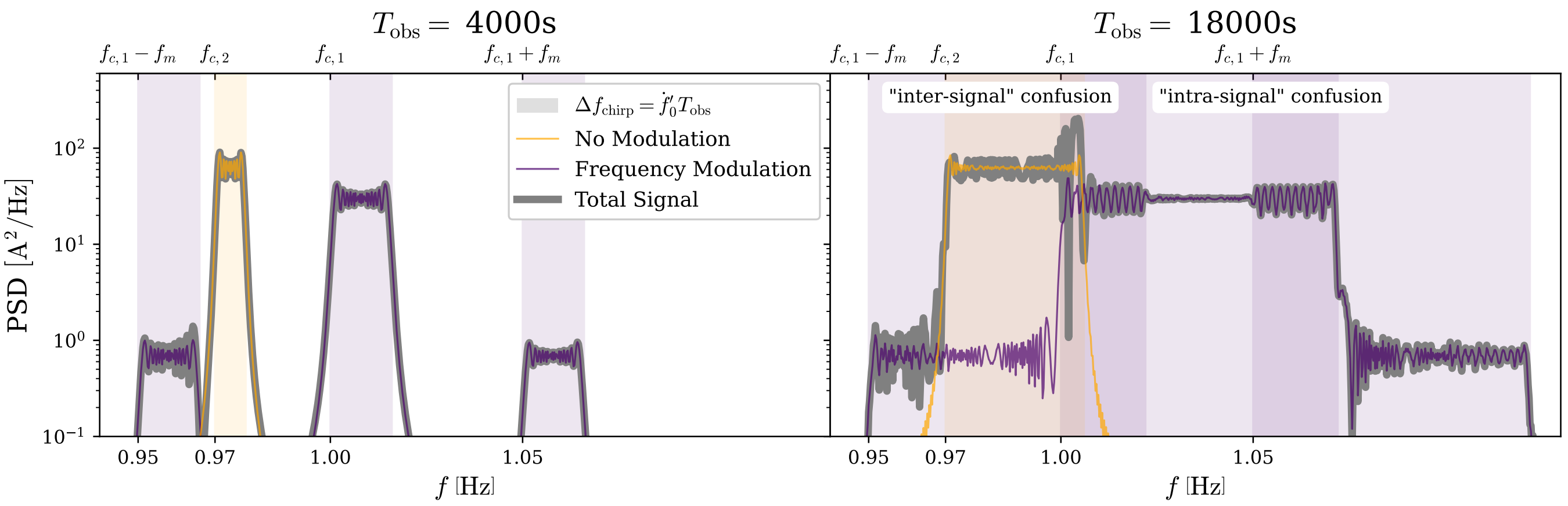

**Figure C2.** Intersignal vs. intrasignal confusion in chirping signals. When observing the power spectrum of a signal that is evolving in frequency (chirping), over time the signal may mix with other nearby signals (intersignal confusion), or mix with itself (intrasignal confusion). This example shows two signals, a nonmodulated signal and an FM signal. In this example, the nonmodulated signal is evolving at half the rate of the FM signal. Left: these two signals ($f_{c,1}$ and $f_{c,2}$) are very close to each other in frequency space (the signal with no modulation is within the sidebands of the FM signal). Right: for longer observation times, the frequency chirping causes the signals to grow and span a wider spectrum of frequencies $\Delta f_{\rm chirp}$. Where the two signals overlap with each other, we label it as "intersignal" confusion, and where the FM signal overlaps with itself (i.e., the carrier and sideband messages mix), we label it as "intrasignal confusion." Mathematically, this intrasignal confusion occurs when $\Delta f_{\rm chirp} \geqslant f_m$. Additionally, we point to another important feature of the FM signal—its sideband messages chirp at the same rate as the carrier signal.

Similarly, the PSD evaluated at the representative peaks $\omega = \omega_c + n\omega_m + \frac{1}{2}\dot{\omega}_0' T$ gives

$$S_M^{(1)}\left(f = f_c + nf_m + \frac{1}{2}\dot{f}_0' T\right) \approx \frac{A^2 J_n^2(\epsilon)}{2|\dot{f}_0'|\ T}. \tag{C10}$$

### C.4. Nonmodulated Signals

Note that if we want the nonmodulated versions of each of these signals, then that corresponds to the limit that $\epsilon \to 0$. In this case, since $J_0(0) = 1$, but $J_{n\neq 0}(0) = 0$, all terms in the summations of the Fourier transform Equations (C3), (C6), and (C9) vanish except the $n = 0$ term. This also removes the sideband peaks and dependence on the Bessel function from the PSDs (Equations (C4), (C7), and (C10)). Examples of nonmodulated signals (both monochromatic and chirping) are shown in panel (c) (top row) of Figure C1.

### C.5. GW Source + Exoplanet Signal

Finally, we can now use the results of the previous sections to write down the Fourier transform of the GW strain (Equation (11)). If we assume constant amplitudes and write

$$\begin{aligned} H_+(t) &= A_1 \cos(\Phi(t)) + A_2 \cos\left(\Phi(t) + \frac{\pi}{2}\right) \\ &\equiv M_1(t) + M_2(t), \\ \text{where} \quad A_1 &\equiv -2A\frac{(1+\cos^2(\iota))}{2}\cos(2\psi), \\ A_2 &\equiv -2A\cos(\iota)\sin(2\psi), \end{aligned} \tag{C11}$$

then the Fourier transform is

$$\begin{aligned} \tilde{H}_+(f) &= \tilde{M}_1(f;\, \phi_0', \dot{\omega}_0') + \tilde{M}_2\left(f;\, \phi_0' + \frac{\pi}{2}, \dot{\omega}_0'\right), \\ \text{where} \quad \phi_0' &\equiv \phi_0 - \epsilon \sin(\phi_p), \\ \text{where} \quad \dot{\omega}_0' &\equiv 2\dot{\omega}_0. \end{aligned} \tag{C12}$$

Notice that to properly use the results derived in the previous sections, the initial phase parameter needs to be redefined to $\phi_0'$ to account for the additional $\epsilon\sin(\phi_p)$ term in Equation (6). The frequency chirp term also must be set to $2\dot{\omega}_0 = 2\pi\dot{f}_{\mathrm{GW},0}$, as we would expect. Depending on which signal model we want, we can now use Equation (C3), Equation (C6), or Equation (C9) for $\tilde{M}_1$ and $\tilde{M}_2$.

## ORCID iDs

Casey McGrath https://orcid.org/0000-0002-6155-3501